\documentclass[twocolumn]{openjournal}

\usepackage{natbib}
\usepackage{graphicx}	
\usepackage{amsmath}	
\usepackage{amssymb}	
\usepackage{xcolor}
\usepackage[normalem]{ulem} 
\usepackage{newtxtext,newtxmath}
\usepackage[breaklinks=true,
hyperindex=true,
colorlinks=true,
citecolor=blue]{hyperref}

\newcommand{\Msun}{$\mathrm{M}_{\odot}$}

\begin{document}

\title{The role of turbulence in setting the phase of the ISM and implications for the star formation rate}

\author{Tine Colman$^{* \ 1,2}$}
\author{Patrick Hennebelle$^{1}$}
\author{No\'e Brucy$^{3,2}$}
\author{Pierre Dumond$^{2}$}
\author{Philipp Girichidis$^{3}$}
\author{Simon~C.~O.~Glover$^{3}$}
\author{Ralf S.\ Klessen$^{3,4}$}
\author{Marc-Antoine Miville-Desch\^enes$^{5}$}
\author{Sergio Molinari$^{6}$}
\author{Rowan Smith$^{7,8}$}
\author{Juan~D.~Soler$^{6}$}
\author{Leonardo Testi$^{9,10}$}
\author{Alessio Traficante$^{6}$}

\vspace{1 cm}

\email{$^*$email: tine.colman@cnrs.fr}

\affiliation{$^{1}$ AIM, CEA, CNRS, Université Paris-Saclay, Université Paris Diderot, Sorbonne Paris Cité, 91191 Gif-sur-Yvette, France}
\affiliation{$^{2}$ ENS de Lyon, CRAL UMR5574, Universite Claude Bernard Lyon 1, CNRS, Lyon 69007, France}
\affiliation{$^{3}$ Universit\"{a}t Heidelberg, Zentrum f\"{u}r Astronomie, Institut f\"{u}r Theoretische Astrophysik, Albert-Ueberle-Str. 2, 69120 Heidelberg, Germany}
\affiliation{$^{4}$ Universit\"{a}t Heidelberg, Interdisziplin\"{a}res Zentrum für Wissenschaftliches Rechnen, Im Neuenheimer Feld 205, 69120 Heidelberg, Germany}
\affiliation{$^{5}$ Laboratoire de Physique de l'École Normale Supérieure, ENS, Université PSL, CNRS, Sorbonne Université, Université de Paris, F-75005 Paris, France}
\affiliation{$^{6}$ INAF -- Istituto di Astrofisica e Planetologia Spaziali, via Fosso del Cavaliere 100, 00133 Roma, Italy}
\affiliation{$^{7}$ Jodrell Bank Centre for Astrophysics, Department of Physics and Astronomy, University of Manchester, Oxford Road, Manchester M13 9PL, UK}
\affiliation{$^{8}$ SUPA School of Physics and Astronomy, University of St Andrews, North Haugh, St Andrews, Fife, KY17 9SS, UK}
\affiliation{$^{9}$ Alma Mater Studiorum Università di Bologna, Dipartimento di Fisica e Astronomia (DIFA), Via Gobetti 93/2, I-40129, Bologna,
Italy}
\affiliation{$^{10}$ INAF-Osservatorio Astrofisico di Arcetri, Largo E. Fermi 5, I-50125, Firenze, Italy}


\begin{abstract}
What regulates star formation in different regions of the Galaxy is still debated and especially the role of turbulence is not fully understood.
In this work, we explore the link between star formation, turbulence and the thermal state of the multi-phase interstellar medium (ISM).
We analyse a suite of stratified box simulations modelling a realistic ISM that aims to probe environments similar to those found in the Milky Way.
Turbulence is injected through stellar feedback and an external large-scale driving force.
We find that star formation can be either boosted or reduced when increasing the external driving strength, depending on the environment.
When the density is sufficiently high or the initial UV background weak, warm neutral gas naturally transitions to the cold phase, leading to high cold neutral medium (CNM) fractions of around 30 -- 40\%.
Under these conditions, excessive large-scale driving leads to a slight reduction of the CNM fraction and an increase in the amount of gas that is thermally unstable.
What limits the star formation in this regime is a reduced fraction of dense gas due to additional turbulent support against collapse.
For low density regions subject to significant external UV background, overdensities in which cooling is efficient are much rarer and we find
that star formation is regulated by the formation of cold gas.
In such cases, turbulence can significantly boost star formation by compressing gas in shocks and increasing the CNM fraction dramatically.
In our simulations we see an increase from almost no CNM to up to a fraction of 15 \% when including external turbulence driving; leading to an associated increase in the star formation rate.
We provide a model to quantify this behaviour and predict the CNM fraction by combining the standard ISM cooling/heating model with the density PDF generated by turbulence.
The change in the dominant limiting process for star formation between low-density /externally heated and intermediate-density /feedback heated environments could provides a natural explanation for the observed break in the Kennicutt-Schmidt relation around column densities of 9\,\Msun\, pc$^{-2}$.
\end{abstract}


\maketitle


\section{Introduction}

The interstellar medium (ISM) consists mainly of hydrogen gas which occurs in ionised (HII), atomic (HI) and molecular (H$_2$) form.
It has been established that HI itself has two thermal phases that coexist at the same pressure: the warm neutral medium (WNM) with temperatures of a few thousand Kelvin and the cold neutral medium (CNM) with temperatures below 250 K \citep{Field_et_al1969, Wolfire_et_al1995, Wolfire_et_al2003, McClure-Griffiths_et_al2023}.
Gas in one of these two regimes is thermally stable, meaning that a small perturbation cannot change its temperature dramatically.
On the other hand, gas with a temperature in the range between the CNM and WNM temperatures is thermally unstable and could undergo rapid transition to either the WNM or CNM state. This regime is called the unstable or lukewarm neutral medium (UNM or LNM).

Gas is the main raw material from which stars form.
As such, relations exist between the gas reservoir and the star formation rate (SFR), which can be averaged over full galaxies or sub-regions within, or measured for individual star-forming clouds.
\citet{Schmidt1959} proposed a power-law relation between the gas density and the SFR.
The relation between the surface density, a quantity which is more directly observable especially for extra-galactic measurements, and SFR is commonly referred to as the Kennicutt-Schmidt (KS) relation \citep{Kennicutt1998, Kennicutt&Evans2012}.
On average, it has the form SFR $\propto \Sigma_\mathrm{HI+H2}^{1.4}$ above 9 \Msun pc$^{-2}$, where the relation exhibits a break, and below which it shows steeper slopes and more scatter \citep{Kennicutt1998, Bigiel_et_al2008}.

The relation becomes even more direct when considering only the molecular gas.
 \citet{Bigiel_et_al2008} have established a tight linear relation $\Sigma_{\mathrm{H}_2} \propto \Sigma_\mathrm{SFR}$, where H$_2$ was estimated from CO emission \cite[see also][]{Pessa2021,Sun2023}.
In contrast, the relation between the SFR and the HI surface density
is found to vary from galaxy to galaxy \citep{Bigiel_et_al2008, Leroy_et_al2008, Shetty2014}.
Since only cold gas is relevant for the formation of stars \citep{Krumholz_et_al2009, Klessen2016}, regardless of whether it is in atomic or molecular form \citep{Glover2012}, we would not expect a universal relation of the SFR with the total amount of HI if the CNM fraction varies with environment.

Decomposition of HI into WNM and CNM is a difficult task and so far observational results are limited to the Milky Way or other Local Group galaxies \citep{Dickey1979, Dickey_et_al2000, HI4PI2016, Murray_et_al2018}. 
In their recent review, \citet{McClure-Griffiths_et_al2023} compiled an overview of CNM fraction measurements in the Galaxy.
A large variety is recovered, with values ranging from zero up to 80\% or more. 
Several trends with environment appear.
\citet{Heiles_Troland2003} found a CNM fraction that on average increases with total HI column density up to $\Sigma_\mathrm{HI} \approx 9$ \Msun pc$^{-2}$, which marks the threshold between HI and H$_2$ dominated gas at solar metallicity \citep{Bigiel_et_al2008, Leroy_et_al2008}.
Observations near molecular clouds typically show relatively high CNM fractions, highlighting the important connection between these two gas phases.
Gas at high Galactic latitudes is found to have lower CNM fractions \citep{Kalberla_Kerp2009, Marchal_Miville-Deschenes2021}.

The star formation process in galaxies can thus be thought of as a sequence of steps: the condensation of CNM out of WNM followed by the possible formation of a molecular cloud out of this CNM, in which then finally dense collapsing cores form and produce stars \citep{Schruba_et_al2011, Smith_at_al2023}.
Star formation can potentially be regulated through either one of these processes and the dominant limiting process is likely dependent on environment.

We know that turbulence plays a key role in star formation \cite[e.g.][]{MacLow2004,McKee2007}.
Observations show that molecular clouds are turbulent and its importance is now widely accepted \citep{Roman-Duval_et_al2010, Hennebelle_Falgarone2012, Mivilledeschenes_et_al2017}.
Analytical theories and simulations show that turbulence interacts with gravity in two ways.
On the one hand, it widens the density PDF, creating more overdensities that may become gravitationally unstable, which is favourable for star formation. 
On the other hand, it provides support against gravitational collapse in the form of turbulent velocity dispersion, which reduces the rate at which dense gas is transformed into stars.
These principles are the basis of the theory of turbulent fragmentation \citep{Krumholz_McKee2005, Padoan_Nordlund2011, Hennebelle_Chabrier2011, Federrath2012, Hennebelle_et_al2024}.

It is thought that the turbulence in molecular clouds is inherited from larger scales during their formation process \citep{Heyer2004,Brunt_et_al2009, Klessen2010}.
What exactly drives the turbulence on these scales is still debated.
In present-day spiral galaxies such as the Milky Way, numerical studies indicate that stellar feedback in the form of supernova (SN) possibly in combination with ionising radiation and winds is the dominant driving mechanism in most parts of the galaxy \citep{Peters_et_al2017, Gatto_et_all2017, Colling_et_al2018, Rathjen_et_al2021}. These studies reproduce levels of star formation in agreement with observations, without needing to invoke additional limiting processes.
Recently, \citet{colman24} investigated simulated cloud properties in different environments and from different numerical setups. They found an overall trend of the turbulent velocity dispersion consistent with large-scale turbulent decay. Furthermore, at small spatial scales -- and small molecular clouds --  nearby SNe leave a noticeable imprint on the velocity dispersion.

The situation may be different in the centre and outskirts of galaxies. \citet{Sun_et_al2020} showed that the velocity dispersion of clouds in the centre of barred spiral galaxies is elevated. The same is true for the central region of our Milky Way \citep{Shetty2012}. The fact that these regions lie below the KS relation \citep{Longmore2013}, leads us to speculate that injection of turbulence from galactic dynamics plays a role here.
Also in galaxies which have high gas densities, such as found at high redshift, large-scale galactic processes may be important drivers of turbulence \citep{Krumholz_et_al2018}.
In their kiloparsec box simulations, \citet{Brucy_et_al2020, Brucy_et_al2023} found that large-scale driving quenches star formation for column densities above 20 \Msun pc$^{-2}$ and that the KS relation cannot be reproduced with stellar feedback alone.



In this work, we study in more detail the effect of large-scale driving in Milky Way-like environments with column densities similar to what is found in the Solar neighbourhood.
The paper is structured as follows.
In Section~\ref{sec:setup}, we describe the simulation setup. A first suite of eight simulations explores two values for the initial column density and various driving strengths. {A constant UV background was used for these run.}
In Section~\ref{sec:SFH}, we analyse and compare the star formation activity in the different simulations in this first set.
We find that, in some cases, driving can enhance star formation rather than reducing it.
In Section~\ref{sec:ISM_phases}, we investigate how turbulence affects the thermal balance in these simulations. Turbulence is found to influence the CNM and LNM fraction.
In Section~\ref{sec:variaUV}, we present a second set of simulations, now with a UV background which is time-dependent and proportional to the global SFR in the box. We repeat the previous analysis and stress the differences with the constant UV case.
The connection to the SFR is explored in Section~\ref{sec:SF_relations}, where we take a closer look at the star formation relations.
We identify two regimes, each with a different limiting process.
In Section~\ref{sec:CNM_model}, we further discuss the generation of CNM through turbulence and construct the prototype of an analytical model to predict the CNM fraction.
Finally, Section~\ref{sec:summary} summarises our findings and concludes the paper.

\section{Simulations}
\label{sec:setup}

\begin{table}
    \centering
    \caption{Overview of the simulations with constant UV background.}
    \begin{tabular}{l l c c r c}
    Group & driving & $\Sigma$\footnote{initial gas column density in \Msun pc$^{-2}$} & $n_0$ \footnote{initial mid-plane density in cm$^{-3}$}  & $f_{\mathrm{rms}}$\footnote{normalisation factor for the turbulence driving strength} & $\sigma$ \footnote{final velocity dispersion in km s$^{-1}$} \\
    \hline \hline
    low-$\Sigma$  & none       & 10.2 & 0.8 & 0    & 2.6\\
    low-$\Sigma$  & very weak  & 10.2 & 0.8 & 1500 & 4.8\\
    low-$\Sigma$  & weak       & 10.2 & 0.8 & 3000 & 7.4\\
    low-$\Sigma$  & medium     & 10.2 & 0.8 & 6000 & 8.3\\
    \hline
    high-$\Sigma$ & none    & 19.1 & 1.5 & 0       & 8.4\\
    high-$\Sigma$ & weak    & 19.1 & 1.5 & 3000    & 9.0\\
    high-$\Sigma$ & medium  & 19.1 & 1.5 & 6000    & 12.1\\
    high-$\Sigma$ & strong  & 19.1 & 1.5 & 24000   & 20.1\\
    \hline
    \end{tabular}
    \label{tab:sims}
\end{table}

As in previous work \citep{Iffrig&Hennebelle2015, Colling_et_al2018, Brucy_et_al2020}, we model a 1\,kpc$^3$ region of a galaxy with realistic ISM physics.
The first set of simulations used for this work are described in \citet{colman22}.
Additionally, we extend this set to include four new runs with lower initial column density, to probe a wider variety of conditions.
What is characteristic about these simulations is that they assume a constant UV background. To gain additional insight, we analysed a second set of simulations with a variable UV field. The details of the first set are described below. In Section~\ref{sec:setup_varUV}, we discuss the differences with the second set.

\subsection{Setup}
The initial condition of the simulations is a smooth, Gaussian density profile perpendicular to the mid-plane, characterised by a mid-plane particle density $n_0$ and a thickness $z_0 = 150$ pc.
Our simulation suite explores two values for the initial density:  $n_0 = 0.8 \, \mathrm{cm}^{-3}$ which is equivalent to a total face-on column density of $\Sigma=10.2 \, \mathrm{M}_\odot \mathrm{pc}^{-2}$, and $n_0 = 1.5 \, \mathrm{cm}^{-3}$ which corresponds to $\Sigma=19.1 \, \mathrm{M}_\odot \mathrm{pc}^{-2}$. We  refer to these sub-sets of models as low-$\Sigma$ and high-$\Sigma$.

We include an initial magnetic field with an orientation along the $x$-axis. The magnetic field strength has a Gaussian distribution in the vertical direction with the same thickness $z_0$ as the gas and a mid-plane value $B_0 = 7.62 \, \mu$G, which is compatible with various estimates for the magnetic field in the Milky Way \citep{Beck2003, Heiles_Crutcher2005, Heiles_Troland2005}.

An initial level of turbulence is introduced by adding a turbulent velocity field with a root mean square velocity dispersion of 5 km s$^{-1}$ and a Kolmogorov power spectrum $E(k) \propto k^{-5/3}$ with random phase.
These fluctuations will kick-start structure formation.
The initial temperature is uniform and set to 5333 K, which is a typical value for the warm neutral medium phase in the ISM.

\subsection{Numerics}

To evolve our simulation in time, we used the adaptive mesh refinement magneto-hydrodynamics code \textsc{ramses} \citep{Teyssier2002, Fromang_et_al2006}, with treatments of the magnetic field in the ideal approximation and radiation using a moment-based method \citep{Rosdahl_et_al2013}.
The simulation box is periodic in the $x$- and $y$-direction and has open boundary conditions in the $z$-direction.
The coarse grid has a resolution of 3.9 pc and is further refined based on a mass criterion up to a maximum resolution of 0.24 pc in the densest regions.
Aside from self-gravity, we also apply an external gravitational potential representing the effects of the stellar disk as prescribed by \cite{Kuijken_Gilmore1989, Joung_MacLow2006}.
We use the ISM cooling/heating model from \cite{Audit_Hennebelle2005} which is based on the work of \citet{Wolfire_et_al2003}, further detailed in one of the next subsections. 
This model produces a multi-phase ISM with a warm and cold phase in our simulations, as demonstrated later on in this paper.

\subsection{Sink particles and stellar feedback}

When the gravitational collapse reaches the resolution limits, we introduce sink particles \citep{Bate_et_al1995, Federrath_et_al2010} according to the recipe prescribed by \cite{Bleuler&Teyssier2014}. 
After their birth, sinks accrete gas according to the threshold accretion scheme: gas which is above the sink formation threshold of $10^4$ cm$^{-3}$ and within the accretion radius of 4 cells will be accreted.
The high-$\Sigma$ simulations have been evolved for 60 Myr during which 2 to 5 $\%$ of the gas has been converted into stars.
The low-$\Sigma$ runs are computationally less demanding and have varying final times, depending on the intensity of their star formation activity.

Each time a sink has accreted 120 \Msun of gas, it forms an individual massive star with a mass between 8 and 120 \Msun, randomly drawn from the Salpeter IMF \citep{Salpeter1955}.
The remainder of the mass in the sink particle is considered to be distributed between low mass stars.
The total mass of a sink is thus equal to the sum of all individual massive stars that formed in this sink, in addition to the mass of an unresolved collection of low-mass stars.

The massive stars are sources of stellar feedback.
They emit ionising radiation from the position of the sink particle in which they formed and explode as a supernova at the end of their lifetime at a random location around the sink  \citep{Rosdahl_et_al2013, Geen_et_al2016,Iffrig&Hennebelle2017,Colling_et_al2018}.
The supernovae inject $4 \times 10^{43}$ g cm s$^{-1}$ momentum and $10^{51}$ erg of thermal energy, but the velocity and temperature of the surrounding gas are limited to 300 km s$^{-1}$ and $10^6$ K respectively to prevent the time step from becoming very small.
 We have tested the impact of this procedure by comparing several statistics of runs with and without these restrictions and find limited differences. A comparison regarding the SFR has been presented 
in \citet{Brucy_et_al2023}. In the 
Appendix~\ref{sn_treatment}, we additionally show the velocity dispersion as a function of 
density. We find that the limitations do not significantly affect the results.
The physical reason is that when very high velocities (typically larger than 300 km s$^{-1}$) are needed, it is because the supernova explodes in a very tenuous medium. Such explosions have very limited impact on the dense gas, as studied for instance 
by \citet{Iffrig&Hennebelle2015}.
For low-mass stars, the UV output is negligible and their lifetimes are much longer than the simulation time.
Due to their computational cost, we do not include stellar winds.
Furthermore, similar numerical studies found that radiation is more important for the regulation of the SFR and that including winds further reduces the SFR by only a factor 2 \citep{Rathjen_et_al2021}.

\subsection{Turbulence driving}

\begin{figure*}
    \centering
    \includegraphics[width=\textwidth]{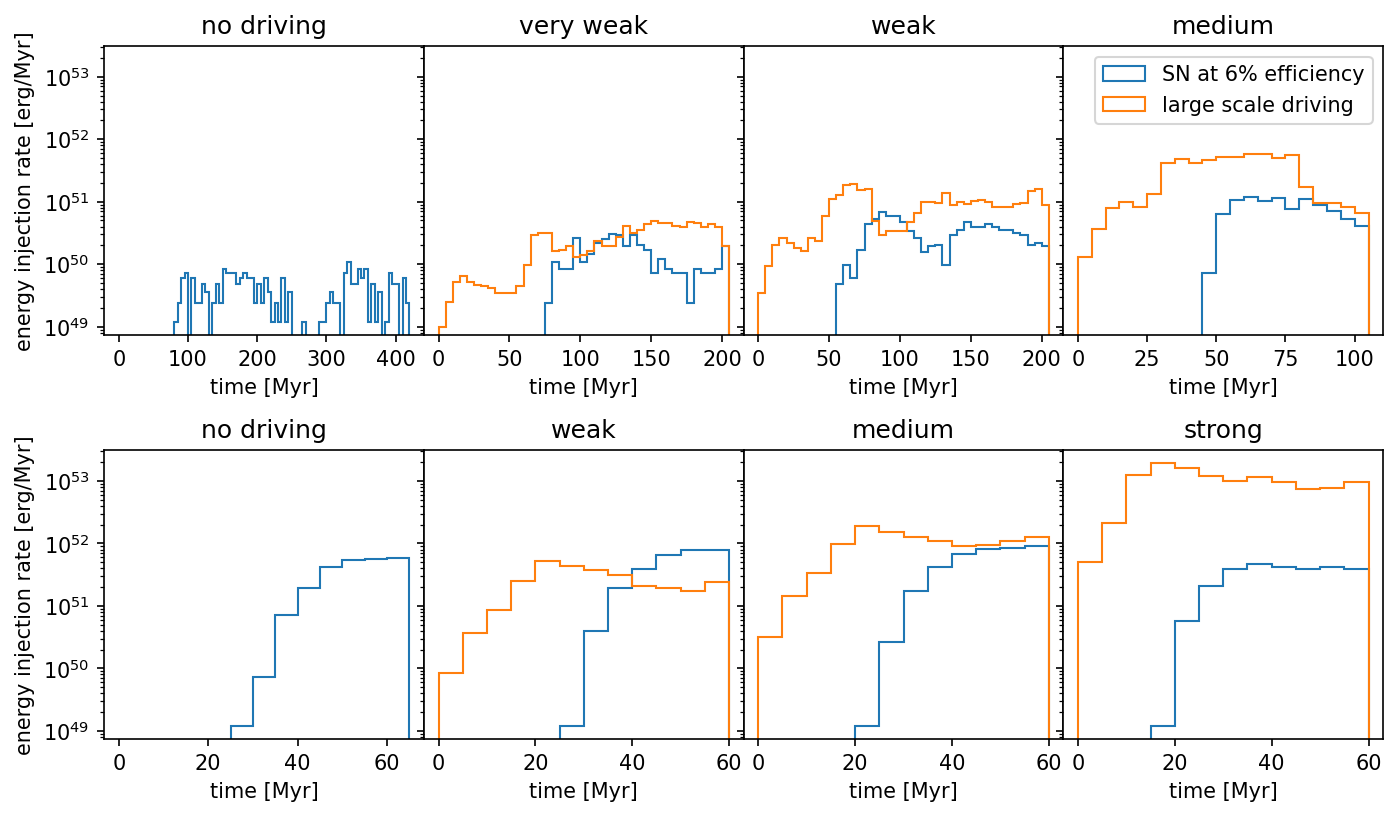}
    \caption{Comparison between the energy injection rate by SNe, assuming a 6\% efficiency, and large-scale driving throughout the simulations averaged over bins of 5 Myr.}
    \label{fig:energies}
\end{figure*}

\begin{figure}
    \centering
    \includegraphics[width=\columnwidth]{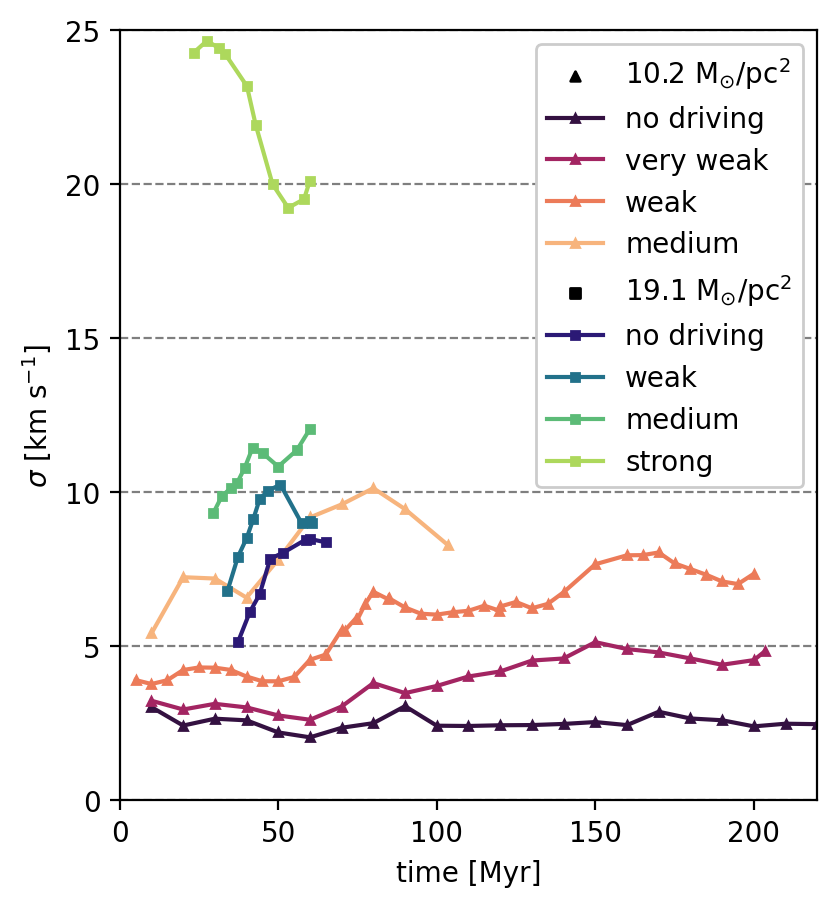}
    \caption{Evolution of the mass-weighted 3D velocity dispersion in the different simulations, measured in a region between 100 pc above and 100 pc below the mid-plane.}
    \label{fig:velocity_dispersion}
\end{figure}

\begin{figure*}
    \centering
    \includegraphics[width=\textwidth]{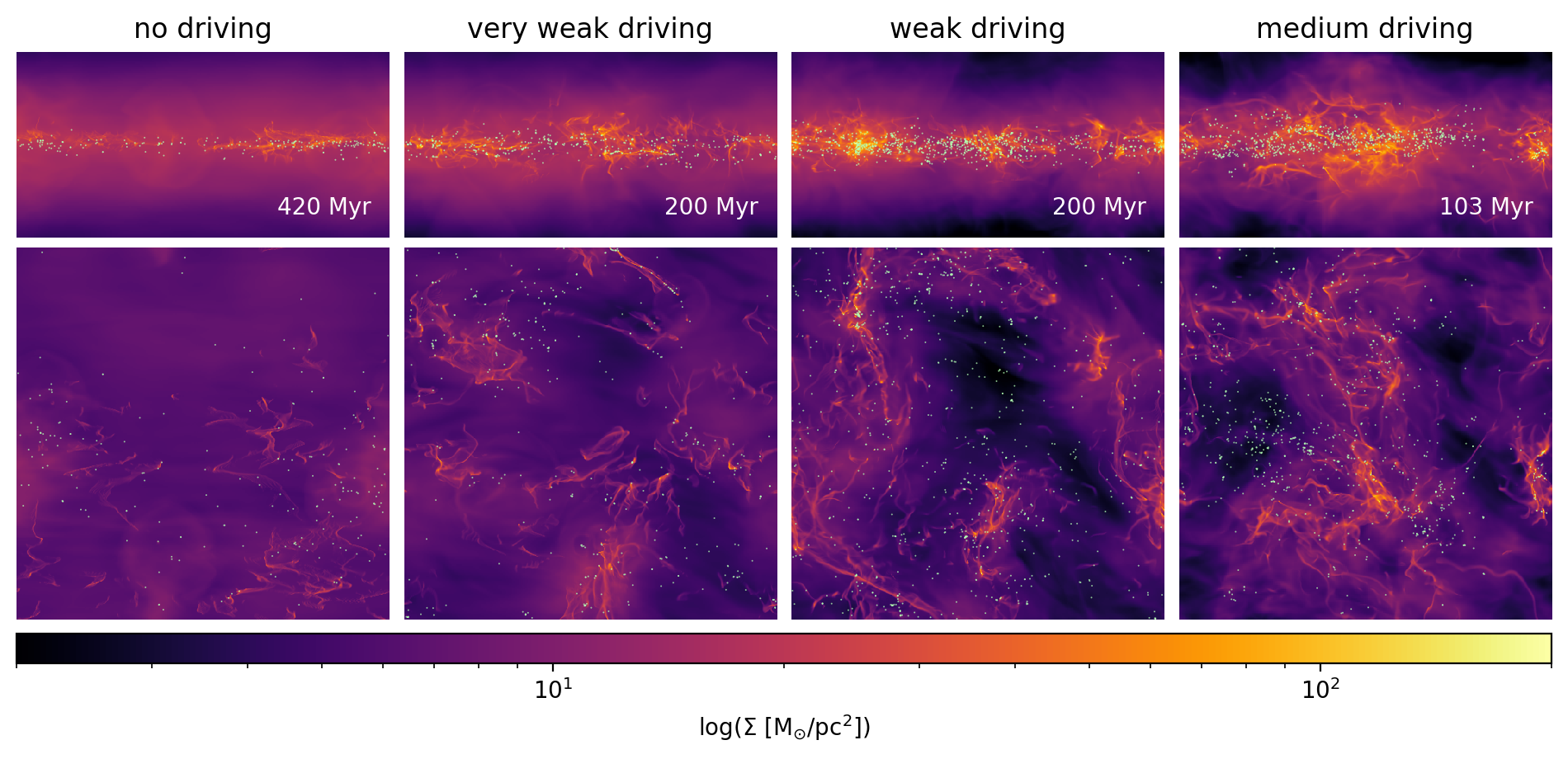}
    \includegraphics[width=\textwidth]{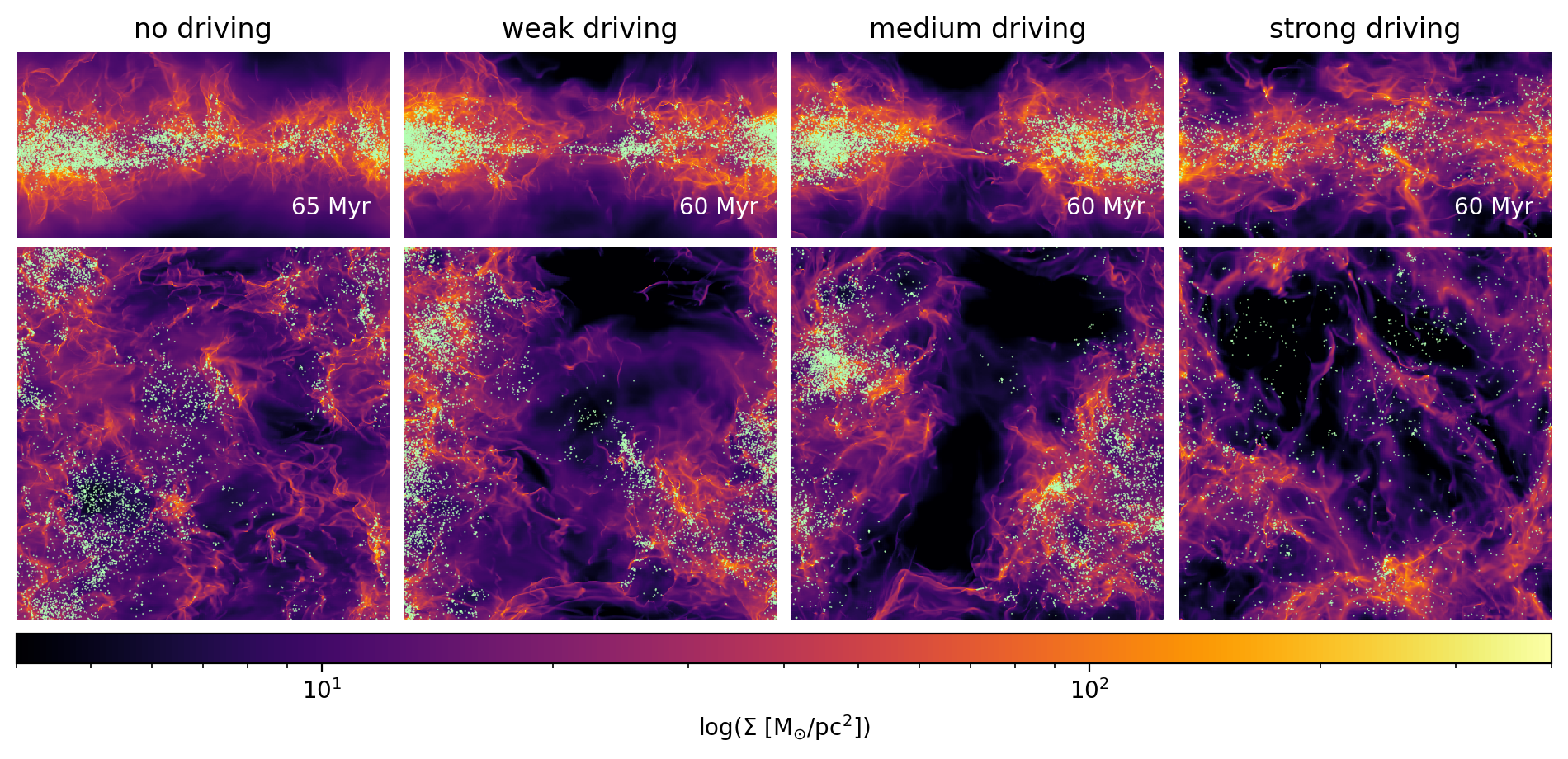}
    \caption{Gas surface density $\Sigma$ for edge-on and face-on projections of the final snapshot of each simulation. The field of view is 1 kpc $\times$ 0.5 kpc and 1 kpc $\times$ 1 kpc respectively, each with a depth of 1 kpc. 
    The green dots indicate the location of the sink particles.  Top: low-$\Sigma$ runs with $\Sigma = 10.2$ \Msun pc$^{-2}$. Bottom: high-$\Sigma$ runs with $\Sigma = 19.1$ \Msun pc$^{-2}$.}
    \label{fig:column_density_maps}
\end{figure*}

In this work, we study the effect of turbulent energy injection from large scales.
This presents a source of turbulence additional to stellar feedback, mimicking the possible effect of galactic dynamical processes that occur on scales larger than the box size, such as spiral arms, a bar, or interactions with neighbouring galaxies.
A detailed prescription of the turbulence driving can be found in \citet{colman22} or \citet{Brucy_et_al2023}.
In short, a bi-dimensional turbulence driving force aligned with the disk plane is added as an additional stochastic external force in the Euler equation.
This driving acts on scales between a full and a third of the box length and is mainly solenoidal, with a solenoidal fraction of 0.75.
We explore several different forcing strengths, ranging from very weak to strong, as well as running reference simulations without external forcing (see Table~\ref{tab:sims}).

In Figure~\ref{fig:energies} we compare the energy injected by large-scale driving to an estimate of the turbulence injection by SNe per unit time for each simulation. We count an energy of 10$^{51}$ erg at an average efficiency of 6\% per SN. Note that this efficiency is difficult to quantify and it depends on the environment in which the SN goes off \citep{Martizzi_et_al2015,Iffrig&Hennebelle2017}.
After an initialisation phase during which the SN rate increases, the feedback self-regulated to a fairly stable rate which is mostly dependent on the column density of the simulation.
Figure~\ref{fig:energies} serves as a guide for interpreting the relevance of both driving mechanisms in the remainder of this paper.

Different forcing strengths lead to different final global velocity dispersions.
In Figure~\ref{fig:velocity_dispersion}, we show the evolution of the velocity dispersion in the different simulations, measured in the disk.
This quantity reflects the differences in energy injection rate from Figure~\ref{fig:energies}.
We note that the values obtained in some of the low-$\Sigma$ runs are only around 2-3 km s$^{-1}$, which is significantly lower than what is usually inferred in simulations and observations.
This is due to the low SN rate in these runs. We will come back to this in Section~\ref{sec:varUV_velocity_dispersion}.
For the other simulations, our measured velocity dispersions are very similar to what is reported in other work, for instance \citet{girichidis2016} and \citet{kimchang2023} (their Fig. 19 and Table 5 respectively).


From Figures~\ref{fig:energies} and~\ref{fig:velocity_dispersion}, we can estimate the length of the initialisation phase and after how much time the initial conditions have been washed out.
For the high-$\Sigma$ runs, the SN rate is converged after 50 Myr or less. The stronger the driving, the shorter the initialisation phase.
In the low-$\Sigma$ runs, the evolution proceeds slower especially when there is little turbulence.

\subsection{Cooling and heating model}
\label{sec:cooling_heating_model}

What sets the thermal balance in the ISM is a topic that has been well-studied in the past decades.
\citet{Wolfire_et_al1995} summarised which processes are relevant and provide a model for the solar neighbourhood, which was later extended for other Galactic radii \citep{Wolfire_et_al2003}.
More recent improvements include a more accurate treatment for low-metallicity environments \citep{Glover_Clark2014, Bialy_Sternberg2019}.
In the context of this work, we use the ISM cooling/heating model from \cite{Audit_Hennebelle2005}. We assume solar metallicity and solar neighborhood abundances.
We briefly review the dominant processes included in this model.

The dominant heating process is photoelectric heating from small dust grains and PAHs, that is the emission of electrons from the grains after exposure to the interstellar radiation field.
While UV radiation can free electrons in grains of all sizes, these electrons can escape more easily from small grains.
Consequently, the photoelectric heating rate depends on the amount of small grains available and the strength of the far UV field $G_0$.
\citet{Bakes_Tielens1994} provide an analytical model as a function of $G_0$, the temperature $T$ and the electron density $n_e$, assuming a fixed grain size distribution.
Even though the photoionisation of hydrogen is followed using Ramses-RT, the electron density used in the cooling model is calculated using the formula provided by \citet{Wolfire_et_al2003}.
Corrections for photoionisation in the WNM are made based on the work of \citet{Ferland2003}, as is done in \citet{Geen_et_al2016}.
A constant uniform UV background with a strength equal to 0.6 Habing units is assumed \citep{Habing1969}.
In reality, we expect the UV field to vary locally and to be proportional to the local SFR \citep{Ostriker_et_al2010}. 
The advantage of keeping the UV background constant and uniform is that it makes it easy to disentangle between the dynamical effects, induced by the stellar feedback and 
turbulent driving, and the thermal effect induced by UV heating.
In Section~\ref{sec:variaUV}, we will explore a second set of simulations where the UV heating is time-dependent and proportional to the star formation rate (but still uniform spatially).




The recombination of  electrons onto small grains and PAHs \cite[e.g.][]{Draine2003},
which leads to the emission of a photon, is one of  the main coolants at temperatures of 8000 K and above. Together with collisional excitation of the hydrogen Ly$\alpha$ line, it keeps the temperature of the WNM roughly constant as a function of density.
For temperatures below 8000 K, collisional excitation of the [CII] and [OI] fine-structure lines provides the main cooling mechanism. This implies a dependence on the abundances of carbon and oxygen in gaseous form, which are treated as constant in our simulations.

Heating sources of lesser importance are cosmic rays and soft X-rays. In this work, they are represented by a constant uniform background, but we can expect energetic events such as supernovae to boost the importance of these heating mechanisms in specific regions.

The combination of these processes sets the relation between temperature and density in the ISM.
It depends on the strength of the UV field, the presence of small dust grains and PAHs, the abundances of C$^{+}$, O and electrons.
In reality, these can vary with the environment and change the density at which the transition from WNM to CNM occurs (or whether there even exists a pressure range which allows for a two-phase medium).
Future simulations will include a more self-consistent coupling of cooling/heating to the propagation of radiation and ionisation treated by \textsc{ramses-rt}.

\begin{figure*}
    \centering
    \includegraphics[width=\textwidth]{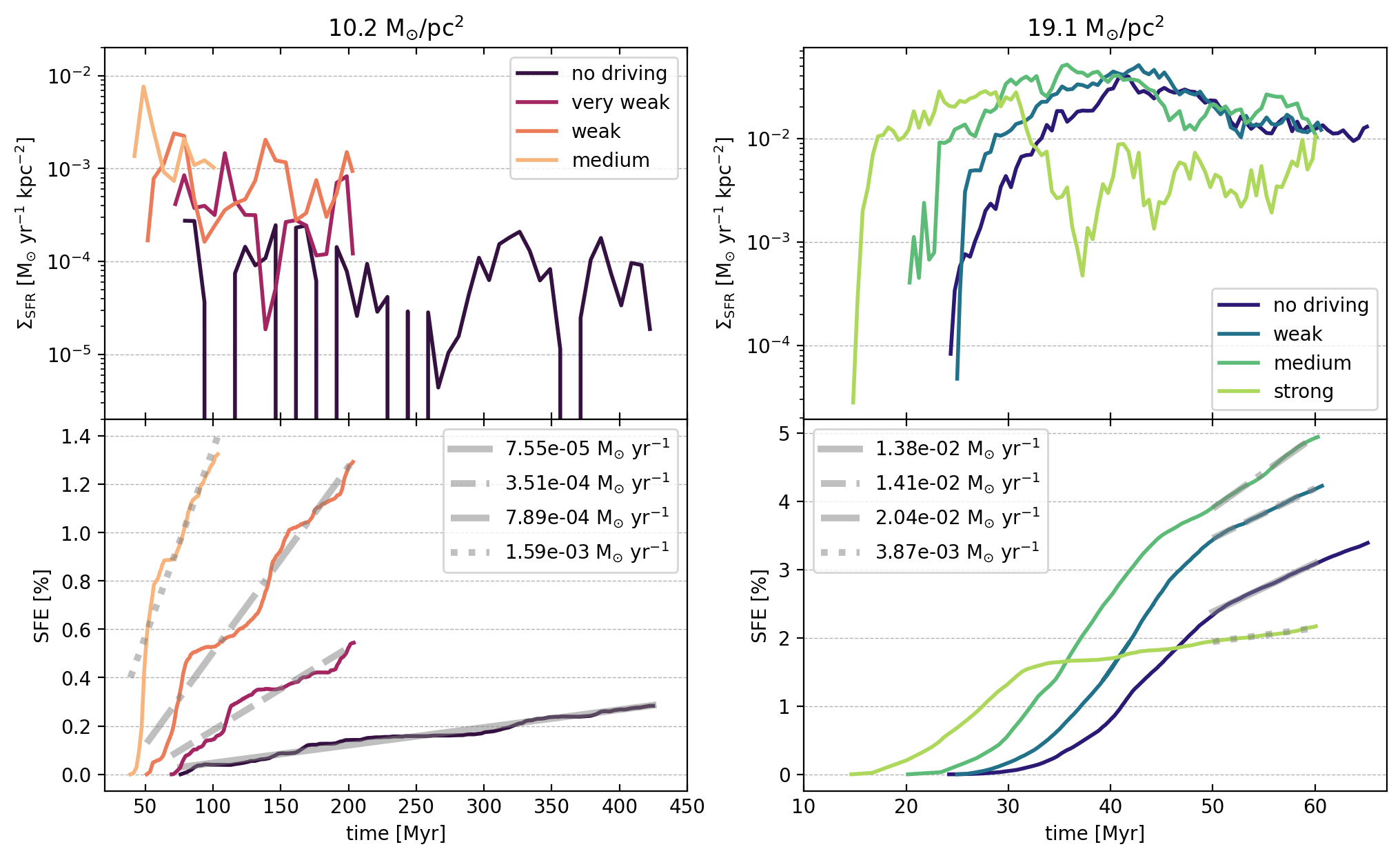}
    \caption{The star formation history in each simulation. Left: low-$\Sigma$. Right: high-$\Sigma$. Top: evolution of the star formation rate, smoothed over a time scale of 7.5 Myr for low-$\Sigma$ and 0.5 Myr for high-$\Sigma$. Bottom: evolution of star formation efficiency.
    The grey lines show the fits to estimate the global SFR of the simulation. The labels indicate the average SFR value found from these fits. 
    }
    \label{fig:sfh}
\end{figure*}

\subsection{Simulation suite with constant UV field}

The parameters of the first set of simulations are summarised in Tables~\ref{tab:sims} and the final snapshot of each simulation is displayed in Figure~\ref{fig:column_density_maps}.
We can see a wide variety in terms of structure.
The low-$\Sigma$ runs (top panels), especially the ones with not too weak driving, have a prominent diffuse gas component with a thin disk of denser material in which stars form.
The high-$\Sigma$ runs (bottom panels) are more filamentary.
Increasing the driving strength leads to more large-scale structure, seen as large overdensities and voids \citep{colman22}.

The various simulations can be compared to different regions in the Milky Way.
The low-$\Sigma$ runs could be placed in the Solar neighbourhood, or more generally at a Galactocentric radius between about 5 and 12 where the average HI column density of the Milky Way is around 10 \Msun pc$^{-2}$. Since the chosen background UV field is lower than the Solar neighbourhood value, an inter-arm region or region beyond the solar galactocentric radius is likely more appropriate.
The high-$\Sigma$ runs have column densities that are somewhat higher than what is found in most parts of the Milky Way disk. These runs may be compared to regions that are slightly closer to the centre of the Galaxy. In this case, strong large-scale turbulence driving may mimic the effect of the central bar.

\section{Star formation in the simulations}
\label{sec:SFH}

\subsection{Star formation history}

Using sink particles, we trace the star formation activity throughout the simulations.
In the context of this work, sinks represent small star clusters rather than individual stars, and consists of a number of massive stars in addition to an unresolved collection of low-mass stars.
Figure~\ref{fig:sfh} displays the star formation history of each simulation.
The top panels show the variation of the SFR in time, which exhibits large fluctuations.
The bottom panel shows the evolution of the SFE, defined as the amount of gas which is in sinks compared to the total mass available at the beginning of the simulation.
The general behaviour is different for the low-$\Sigma$ runs (10.2 $\mathrm{M}_\odot \mathrm{pc}^{-2}$, left panels) compared to the high-$\Sigma$ runs (19.1 $\mathrm{M}_\odot \mathrm{pc}^{-2}$, right panels) and we see two regimes emerging.

For the low-$\Sigma$ simulations, large-scale turbulence driving boosts star formation.
Both the SFR and SFE increases significantly with increasing driving strength.
Without driving there is overall little star formation activity and there are epochs where there is no star formation at all.
Consequentially, there is also little turbulence injection by stellar feedback (Figure~\ref{fig:energies}).
In all cases, the evolution consists of a sequence of alternating bursts and quiescent phases, resulting in a characteristic staircase pattern in the SFE evolution.
Even in the no-driving case we recover this pattern, though the amplitude of the bursts is much lower.
Increasing the driving strength increases the intensity of the bursts and shortens the length of quiescent phases.

In the high-$\Sigma$ runs, we see that weak and medium driving boost the overall SFE compared to the case without external driving. The SFR at the end of the simulation is similar between the no-driving, weak and medium driving cases.
For strong driving however, we see a significant decrease in both SFR and SFE.
These runs demonstrate the duality of turbulence: adding a weak to moderate amount of turbulence slightly enhances star formation, but cranking up the driving strength too much leads to quenching.
We also note that star formation starts earlier when we increase the driving strength.
Since we start from a smooth density profile, cold gas structures have to be generated from scratch by the velocity perturbations. The additional external turbulence helps to kick-start cloud formation.
This is followed by a first phase of accelerating star formation, during which stellar feedback establishes its role in the SFR regulation.
Note that this phase is shorter for the strong driving case.
After the initialisation period, the evolution of the SFE becomes linear, indicating a roughly constant SFR, as opposed to the bursty nature seen at lower $\Sigma$.
Possibly, a similar staircase pattern would emerge if the simulations were evolved for a longer time.

The qualitative findings described here for modest column densities are different from the results obtained for columns densities above 20 \Msun $\mathrm{pc}^{-2}$, where large-scale driving was found to always reduces the SFR \citep{Brucy_et_al2020, Brucy_et_al2023}. This indicates that the relative importance of the processes involved in star formation varies with the column density of the region.

\subsection{Global SFR}

We can obtain an overall SFR for each simulation by fitting a linear function to the SFE evolution in the appropriate time frame.
For high-$\Sigma$, we choose the range between 50 and 60\,Myr where all simulations are in the linearly increasing SFE regime.
We find values of $1.4 \times 10^{-2}$ \Msun yr$^{-1}$ for no-driving and weak driving, $2.0 \times 10^{-2}$ \Msun yr$^{-1}$ for medium driving and $0.39 \times 10^{-2}$ \Msun yr$^{-1}$ for strong driving, i.e. a reduction of roughly a factor 4 compared to the others.
To determine the value for the SFR in the low-$\Sigma$ regime, we simply fit the SFE over the entire time range from the moment the first stars form, averaging over bursts and quiescent phases.
We obtain a value of $7.6 \times 10^{-5}$ \Msun yr$^{-1}$ for the no-driving case, $3.5 \times 10^{-4}$ \Msun yr$^{-1}$ for very weak driving, $7.9 \times 10^{-4}$ \Msun yr$^{-1}$ for weak driving and $1.6 \times 10^{-3}$ \Msun yr$^{-1}$ for medium driving.
Here, the SFR increases with driving strength and the additional turbulence can boost the SFR by more than an order of magnitude.


\section{Thermal state of the gas}
\label{sec:ISM_phases}

In the previous section, we showed that turbulence can either boost or reduce the SFR, depending on the global conditions of the simulation box.
To gain insight into this behaviour, we turn our attention to the thermal state of the two-phase ISM.

\subsection{Gas phases}

\begin{figure}
    \centering
    \includegraphics[width=\columnwidth]{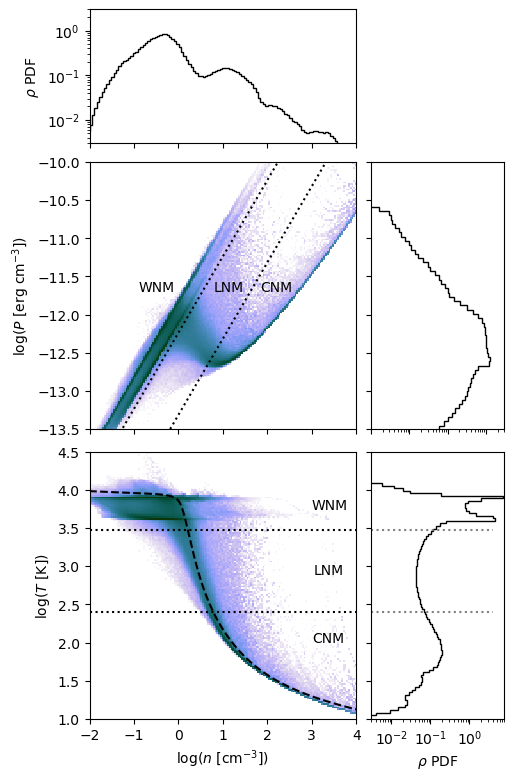}
    \caption{An example of a phase diagram, which show the relation between pressure $P$ or temperature $T$ and number density $n$ in the simulations.
    The colour corresponds to the mass fraction, on a logarithmic scale, with dark being high and light being low. We also show the mass-weighted histograms of each quantity. The dashed black line shows the theoretical cooling model, while the dotted lines indicate the temperature thresholds we adopts for the definition of WNM and CNM in this work.
    }
    \label{fig:phase_diagram}
\end{figure}

In Figure~\ref{fig:phase_diagram}, we show the phase diagram of all the gas (neutral and ionised) in a snapshot of the low-$\Sigma$ simulation with weak driving. For the other simulations, the diagram look qualitatively similar. It shows the relation between density and temperature or pressure, as well as the mass weighted histogram of each of those quantities.
The pressure versus density diagram exhibits the characteristic S-shape with the WNM as the left diagonal branch and the CNM as the right diagonal branch. In between sits the LNM, which connects the two states.
Both the warm and cold medium exist in the same pressure range. The typical ISM pressure is around 3 $\times$ 10$^{-13}$ erg cm$^{-3}$. 

On the density-temperature diagram we overlay the theoretical curve given by the cooling model as a dashed black line.
The WNM has a temperature between 3000 and 10$^4$ K. 
Two distinct WNM branches with each their characteristic temperature are visible: one at $T \approx 4500$ K ($\log T = 3.65$) and the other at $T \approx 8000$ K ($\log T = 3.90$). In the appendix, we show that these can be associated to different levels of ionisation.
The lower temperature branch corresponds to partially ionised gas (between 0.5 \% and 50\%).
This is likely a numerical artefact caused by inconsistencies in the treatment of cooling and ionisation, as noted in Section~\ref{sec:cooling_heating_model}.
From a density of about 1 cm$^{-3}$, the temperature drops as the gas becomes thermally unstable until it reaches CNM temperature of 250 K and below. The characteristic CNM temperature is between 30 and 100 K.

In the density PDF (top panel), the WNM and CNM phases can be identified as distinct bumps around a characteristic density of about 0.5 cm$^{-3}$ and 30 cm$^{-3}$ respectively.
Note however that not all dense gas is cold.
Dense shells around newly formed stars are heated and partially ionised by the UV radiation emitted by those stars. This makes photoelectric heating more efficient and brings the gas to higher temperatures.
Further investigation of this phenomenon requires more careful coupling between photoelectric heating, ionisation and the propagation of the local UV field in the simulations, which is beyond the scope of this work.

In what follows, we will adopt the following definitions for the gas phases.
Neutral gas with a temperature below 250 K is considered to be CNM. This threshold temperature is in agreement with the one used in observations \citep{McClure-Griffiths_et_al2023}.
Note that this definition also includes molecular hydrogen, which is expected to form from cold gas at high densities but whose formation is not tracked explicitly in these simulations.
For the lower bound on the temperature of the WNM we choose 3000 K, a value slightly below what is proposed in \citet{McClure-Griffiths_et_al2023}, to include our lower temperature WNM branch.
The LNM is then everything with temperatures in between these thresholds.
Our boundaries are shown as dotted lines in Figure~\ref{fig:phase_diagram}.


\subsection{Fraction of gas in each phase}

\begin{figure}
    \centering
    \includegraphics[width=\columnwidth]{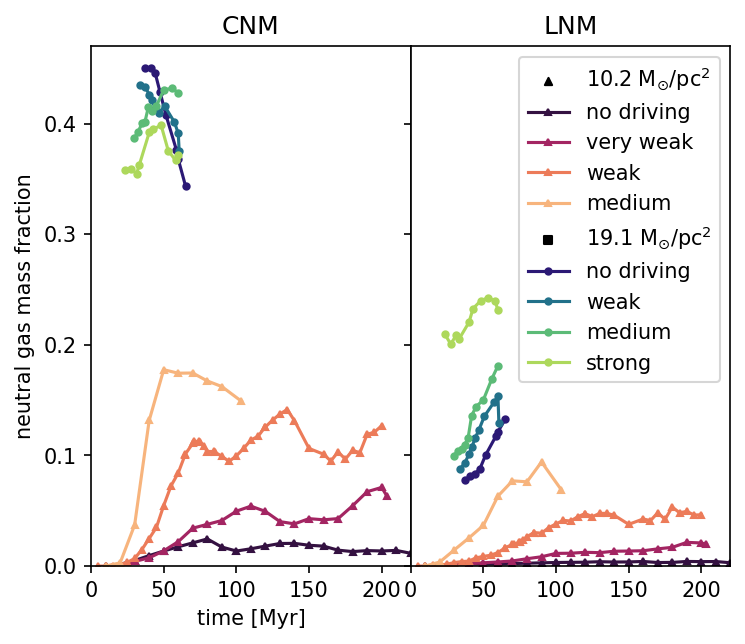}
    \caption{Evolution of the CNM and LNM mass fractions.}
    \label{fig:gas_phase_fractions}
\end{figure}

We quantify the fraction of neutral gas that is in each thermal phase.
In Figure~\ref{fig:gas_phase_fractions}, we show how the CNM and LNM fractions evolve over time in all runs. The WNM fraction is simply $1-f_\mathrm{CNM}-f_\mathrm{LNM}$.

For low-density environments, large-scale turbulence driving significantly promotes the formation of cold material, seen as an increase in both the CNM and LNM fraction. 
Without driving, there is almost no CNM, nor LNM. Turbulence driving thus triggers thermal instability and sets in motion the conversion from WNM to LNM to CNM. The stronger the driving, the larger the increase in CNM fraction. There is up to an order of magnitude more CNM in the driving runs compared to no driving.

For the high-density regions on the other hand, the CNM is less affected by the external driving. The CNM fraction is generally larger than in the low-$\Sigma$ runs, around 40\%. This indicate that the \textit{external} driving is not the dominant mechanism for creating CNM here.
However, possibly turbulence injected by \textit{stellar feedback} plays a similar role. Alternatively, the thermal transition occurred spontaneously due to the larger average density.

We also note important differences in the LNM fraction: driving increases the LNM fraction for all simulations. Notably for the high-$\Sigma$ runs, the ratio of f$_\mathrm{CNM}$/f$_\mathrm{LNM}$ decreases. Driving keeps more gas in the thermally unstable phase, possibly preventing the formation of more CNM or reheating the CNM that was formed before. 
This effect could potentially be used to constrain the level of turbulence in galaxy simulations where the turbulence driving strength in sub-regions is not controlled. Perhaps this relation can also be used to find signatures of turbulence driving in observations, although it remains challenging to observationally quantify the neutral gas phase fractions.

\subsection{Dense gas}
\label{sec:dense_gas}

\begin{figure}
    \centering
    \includegraphics[width=\columnwidth]{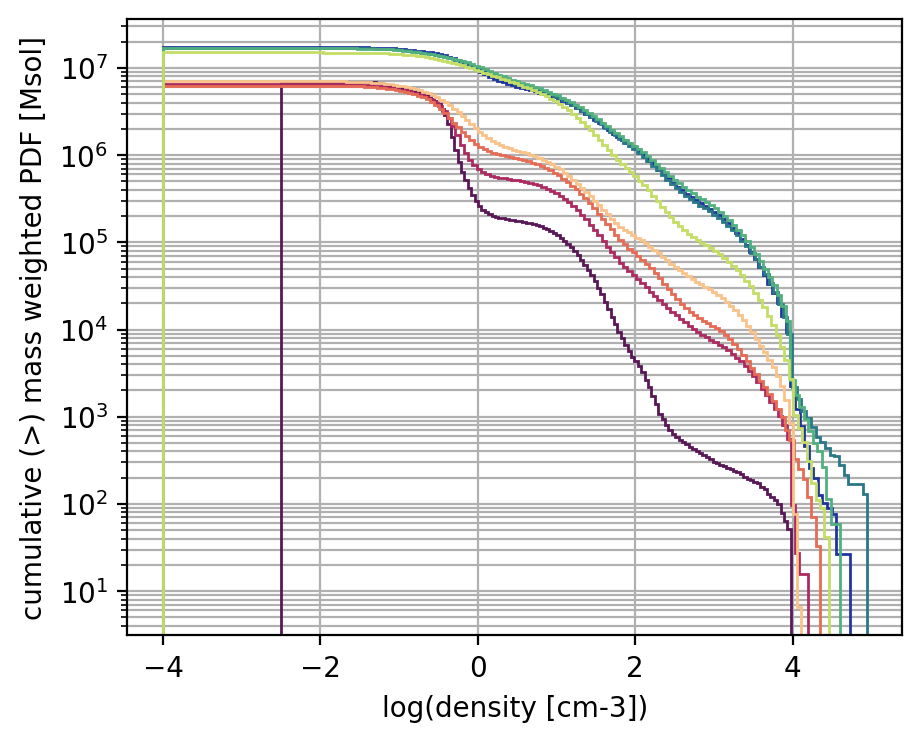}
    \caption{Cumulative distribution function of the density, in absolute gas mass units for a representative snapshot of each simulation.
    The simulations are colour-coded as in Figure~\ref{fig:gas_phase_fractions}.
    }
    \label{fig:cumu_pdf}
\end{figure}

\begin{figure}
    \centering
\includegraphics[width=\columnwidth]{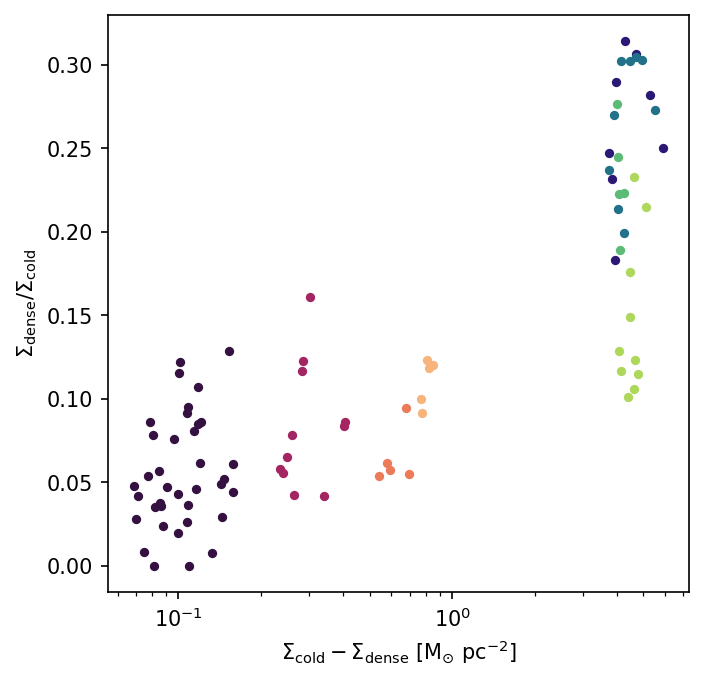}
    \caption{
    The efficiency of the conversion of cold into dense gas, quantified as the fraction of cold gas which has a density higher than 100 $\mathrm{cm}^{-3}$. The x-axis is the column density of the low-density CNM ($n_\mathrm{CNM} < 100 \,\mathrm{cm}^{-3}$). The simulations are colour-coded as in Figure~\ref{fig:gas_phase_fractions}.}
    \label{fig:CNM_vs_H2}
\end{figure}

While stars are found to form mainly in molecular clouds, only the densest parts of these clouds are actively star-forming.
Figure~\ref{fig:cumu_pdf} shows a cumulative distribution function of the density, quantifying how much of the gas is above a certain density. We normalised the curved using the total mass in the simulation.

For the low-$\Sigma$ runs, the large differences in the amount of CNM are again visible here as the curves start to diverge above about 0.5 cm$^{-3}$, the density where the WNM becomes thermally unstable.
This difference propagates to larger densities and only a few percent of the LNM+CNM gas reached densities beyond 100 cm$^{-3}$.

The situation is different for the high-$\Sigma$ runs, which all have a similar cumulative density distribution up to about 10 cm$^{-3}$, at which point the strong driving run starts to deviate from the other three runs. In the case of strong driving, there is significantly less gas with densities above 100 cm$^{-3}$.

We further quantify the efficiency of the formation of dense gas in cold clouds in Figure~\ref{fig:CNM_vs_H2}, where we plot the ratio between the amount of CNM versus the amount of gas with densities greater than 100 cm$^{-3}$.
For a point in the bottom left part of the diagram, the amount of cold material is small and the conversion to high density gas inefficient. If a point is in he top part of the diagram, it means the conversion to dense gas is efficient.
In the low-$\Sigma$ regime the conversion from cold to dense gas occurs at a roughly equal, albeit low, efficiency which is independent of the level of externally driven turbulence.
While the amount of cold HI is similar for all high-$\Sigma$ runs, there is about a factor 2 to 3 less dense gas for the strong driving case.
This brings the efficiency to the levels of the lower column density simulations.
Strong driving hinders the creation of dense self-gravitating clumps.
This is in line with predictions by analytical theories of gravo-turbulent fragmentation \citep{Hennebelle_et_al2024}.

\subsection{A dual role of turbulence on the thermal state?}

It is already established that turbulence has a dual nature with respect to gravity. On the one hand, it generates local overdensities which are the seeds of gravitationally unstable clumps. On the other hand, it provides additional support under the form of turbulent kinetic energy against this exact same gravitational collapse.
This effect is seen in the high-$\Sigma$ runs where strong driving significantly reduces the fraction of dense gas.

Here we demonstrated that turbulence possibly has a second, thermal, dual nature which typically occurs at larger scales. \citet{Hennebelle_Perault1999} already showed that turbulence can create cold condensations in the ISM. Here, we saw that it acts as a perturbing force, promoting phase change and moving the gas away from thermal equilibrium, seen by a consistent increase in LNM fraction (Figure~\ref{fig:gas_phase_fractions}). The duality then lays in the creation versus destruction of CNM.
Our simulation results seem to indicate turbulence has a stronger impact on the WNM-to-CNM conversion than on the reverse conversion. In fact, there is no indication that turbulence is able to bring the gas back into the WNM state, as the WNM fraction decreases and more gas is in the unstable regime.
However, we have to keep in mind that turbulence is not only driven by the external large-scale force that is applied during the simulation. Stellar feedback also drives turbulence.
For the set of high-$\Sigma$ simulations, the turbulent energy injection by SN is either stronger or comparable to that of the large-scale driving, except for the run with strong driving (Figure~\ref{fig:energies}).
Furthermore, the SFR in those three runs is similar, so only in the case of strong driving would we be able to directly link a change in the gas-phase fractions to an excess of turbulence.
If the level of turbulence is a determining factor for CNM formation, we would expect the three other simulations to have similar CNM fractions, as is indeed the case.
Possibly the stellar feedback self-regulated the WNM and CNM fractions.
Controlled simulation of a multi-phase ISM without stellar feedback, could help quantify the dual role of turbulence on the CNM fraction further.
Note that the role of feedback is more complex than just injecting turbulence, as massive stars emit UV radiation which contributes directly to the heating of the ISM.

\subsection{Caveats of a uniform UV background}

The transition point between WNM and CNM depends on the UV radiation field, metallicity and hydrogen ionisation rate \citep{Wolfire_et_al2003}.
While our simulations do track the local UV photon field which is used to determine the ionisation of the gas, this information is not used in the cooling model where a uniform background UV field fixed to the solar neighbourhood value $G_0$ is assumed.
Observations in the Milky-Way and nearby galaxies show that the UV field strength decreases as a function of galactic radius.
The UV field in our low-$\Sigma$ runs may thus be too high compared to real low density regions in galaxies.
A lower UV field implies less heating which would lead to naturally higher CNM fraction.
Furthermore, in the majority of environments we expect the UV field to be directly coupled to the formation of new stars, especially massive ones.
In the next section, we present the result from
simulations where the uniform UV background is proportional to the SFR according to the recipe of \citet{Ostriker_et_al2010}.
We will see that this indeed result in more structure formation and enhanced SFRs compared to the low-$\Sigma$ runs presented in this section.
It remains however unclear how \textit{local} variations in the UV field may affect the overall CNM fraction.
While we demonstrated a clear influence of turbulence under a fixed UV field, in practice regions with conditions similar to those in the low-$\Sigma$ runs may be rare.


\begin{table}
    \centering
    \caption{Overview of the simulations with time-dependent UV background.}
    \begin{tabular}{l c c r c}
    Name & $\Sigma$\footnote{initial gas column density in \Msun pc$^{-2}$} & $n_0$ \footnote{initial mid-plane density in cm$^{-3}$}  & $f_{\mathrm{rms}}$\footnote{normalisation factor for the turbulence driving strength} & $\sigma$ \footnote{final velocity dispersion in km s$^{-1}$} \\
    \hline \hline
    n0.66\_rms00000  & 8.4 & 0.66 & 0    & 11\\
    n0.66\_rms00875  & 8.4 & 0.66 & 875 &  11\\
    n0.66\_rms01750  & 8.4 & 0.66 & 1750 & 11\\
    \hline
    n1\_rms00000  & 12.7 & 1 & 0 & 13\\
    n1\_rms01750 & 12.7 & 1 & 1750       & 13\\
    n1\_rms03500 & 12.7 & 1 & 3500    & 14\\
    n1\_rms07000 & 12.7 & 1 & 7000    & 17\\
    n1\_rms14000 & 12.7 & 1 & 14000   & 20\\
    \hline
    \end{tabular}
    \label{tab:sims_UVvar}
\end{table}

\begin{figure}
    \centering
    \includegraphics[width=\columnwidth]{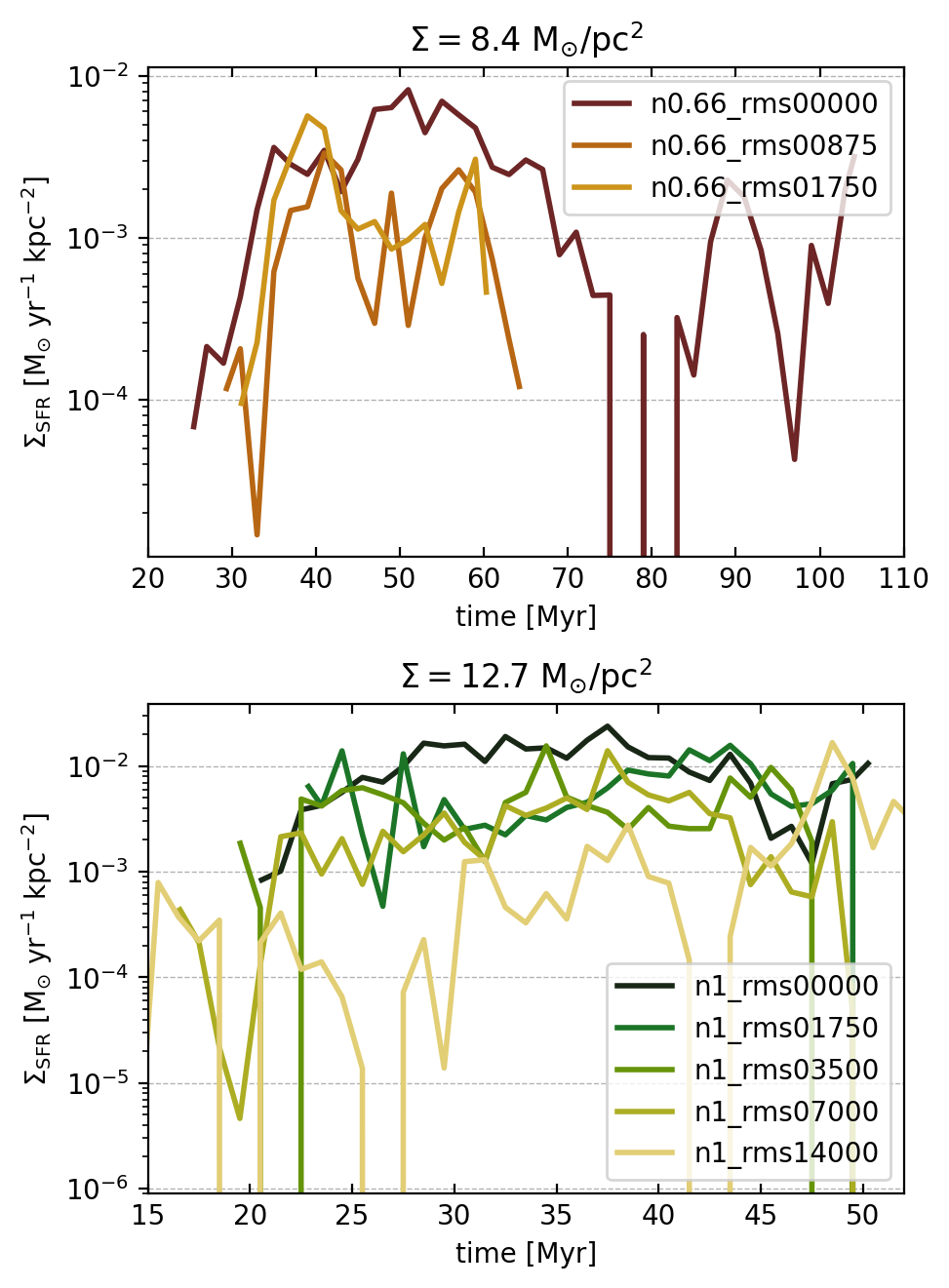}
    \caption{The SFR in the simulations with variable UV background, smoothed over a time scale of 1 to 2 Myr.}
    \label{fig:sfh_ZG}
\end{figure}

\begin{figure}
    \centering
    \includegraphics[width=\columnwidth]{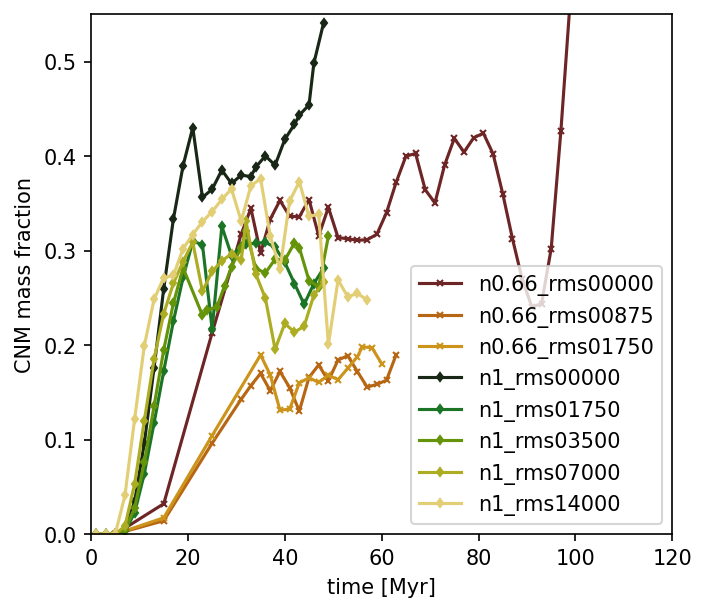}
    \caption{Evolution of the CNM mass fractions in the second set of simulations.}
    \label{fig:gas_phase_fractions_ZG}
\end{figure}

\begin{figure}
    \centering
    \includegraphics[width=\columnwidth]{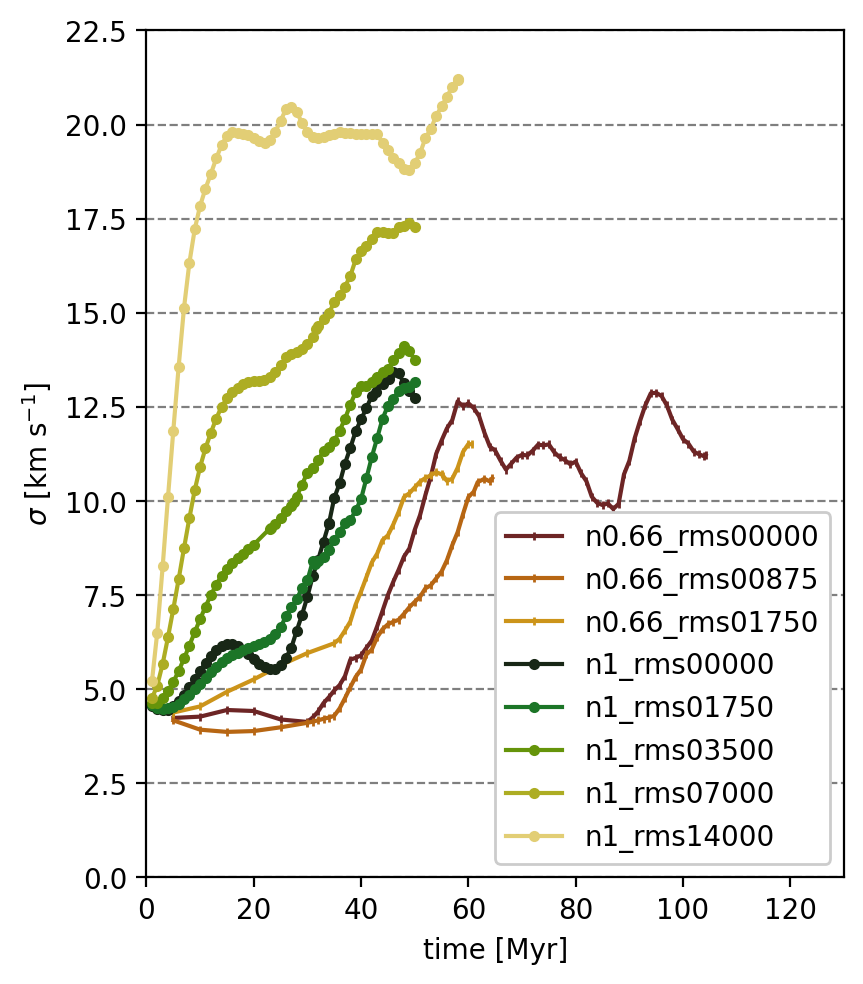}
    \caption{Evolution of the mass-weighted 3D velocity dispersion in the second set of simulations, measured in a region between 100 pc above and 100 pc below the mid-plane.}
    \label{fig:velocity_dispersion_ZG}
\end{figure}


\section{Variable UV background}
\label{sec:variaUV}

As a constant background UV field might not be sufficiently realistic, we now turn our attention to the second set of kpc-box simulations. An overview of their global properties is listed in Table~\ref{tab:sims_UVvar}.

\subsection{Simulation suite with variable UV background}
\label{sec:setup_varUV}

The second set of simulations uses different base parameters than the first set. 
Our goal in this section is to see if we recover similar behaviour as in the previous two sections but for different types of environments, rather than to directly compare the two simulation suites.

As in the first set, we explore two different initial column densities and various external large-scale turbulence driving strengths.
The initial column densities are slightly lower, $\Sigma=8.4 \, \mathrm{M}_\odot\ \mathrm{pc}^{-2}$ and $\Sigma=12.7 \, \mathrm{M}_\odot\ \mathrm{pc}^{-2}$. These values are around the $\Sigma=10 \, \mathrm{M}_\odot\ \mathrm{pc}^{-2}$ were we previously observed a boost in CNM and star formation with increasing external turbulence.
The driving strengths have been adjusted accordingly (see Table~\ref{tab:sims_UVvar}).
The initial magnetic field in the set is also lower, with a value of about 2~$\mu G$.
The spatial resolution is also lower, with a maximum resolution of 1 pc.
The sink formation threshold was adjusted accordingly.

A key difference is the UV background, which is variable in time and proportional to the SFR measured in the computational domain. 
Indeed, we expect the UV field to be proportional to the SFR \citep{Ostriker_et_al2010}.
The mean FUV density relative to the solar neighbourhood value $G^\prime _0$ is
\begin{equation}
    G^\prime _0 = \frac{\Sigma_{SFR}}{\Sigma _{SFR,\odot}} 
= \frac { \Sigma _{SFR} }
{ 2.5 \times 10^{-9} \mathrm{M} _{\odot}  {\rm pc}^{-2} {\rm yr}^{-1}  }.
\label{UV_sfr}
\end{equation}
Additionally, we set a minimum UV field of 0.1 $G_0$.
We use the implementation of \citet{Brucy_et_al2023}.

These simulations track the formation of molecular hydrogen, but because it is not relevant for this study, we do not discuss it here.

\subsection{Star formation rate}

Figure~\ref{fig:sfh_ZG} shows the SFR as a function of time for the eight time-dependent UV background simulations.
The first thing to note is that the simulations without external turbulence driving now actively form stars. The SFR is much higher than the average value of $8\times 10^{-5}$ \Msun\ yr$^{-1}$ we found for the no-driving run with $\Sigma =$ 10.2 \Msun\ pc$^{-2}$ in Section~\ref{sec:SFH}.
Also with driving, the values for the SFR seen in this second set of simulations are somewhat higher than those found in the low-$\Sigma$ runs with constant UV field and higher magnetic field.

In contrast to the results we found in Section~\ref{sec:SFH}, in this second set of simulations, the presence of turbulence driving always leads to a reduced SFR, even when the column density is low. 

\subsection{CNM fraction}

The evolution of the CNM fraction in Figure~\ref{fig:gas_phase_fractions_ZG} shows that it is generally higher than what we found for the low-$\Sigma$ runs in Section~\ref{sec:ISM_phases}.
The UV-field at the beginning of the simulation is set to a very low minimal value. This results in the rapid formation of CNM during the initialisation phase.
When star formation commences after 20 to 30 Myr, the 
CNM fraction has reached values of around 30\% and remains stable afterwards in most cases.

The presence of turbulence now clearly reduces the CNM fraction. For the simulations with $\Sigma=8.4 \, \mathrm{M}_\odot \mathrm{pc}^{-2}$, the CNM fraction of the runs with external driving is only half of that for the no-driving case.
Interestingly, the strength of the driving does not seem to play a role.

\subsection{Velocity dispersion}
\label{sec:varUV_velocity_dispersion}

Figure~\ref{fig:velocity_dispersion_ZG} displays the velocity dispersion as a function of time in the eight simulations of Table~\ref{tab:sims_UVvar}.
A comparison with the runs presented in Figure~\ref{fig:velocity_dispersion} is particularly enlightening. 
Except for the two cases with the strongest driving, the velocity dispersions in this set of simulations reach about 11-13 km s$^{-1}$, irrespective of the forcing strength and initial column density. In contrast, when the UV background is kept constant,  
the runs with 10.2 M$_\odot$ pc$^{-2}$ present significantly lower velocity dispersion ranging from 2 km  s$^{-1}$  when no turbulent driving is applied, to 7 km  s$^{-1}$ when turbulent forcing operates. This stems from the much higher star formation rate that is 
achieved in the simulation with variable UV background. The stellar feedback, in particular the number
of supernova explosions, is one to two orders of magnitude larger than in the case of constant UV background.


\subsection{Constant or variable UV background?}
The results found for simulations with a time-dependent UV background are somewhat different to those with the constant value presented in the previous sections.
Here, the presence of external turbulence driving is always negative, both on the CNM fraction and SFR.

Although a UV background which is local and proportional to the SFR may be assumed to be the norm, there are situations where a constant UV field may be a good approximation.
Let us recall that the mean free path of the UV photons in an ISM with a mean density of 0.3~cm$^{-3}$ is about 1 kpc. This becomes even larger when the metallicity is subsolar.
This implies that for instance in the outskirts of galaxies, the ISM is heated by UV photons which have been emitted at several kpc distance. Accumulated over the galaxy and in a time-span considered by our simulations, the background can then be considered fairly constant.

Moreover, the choice of a constant UV background is a convenient one when comparing to analytical models. 
As will be addressed in Section~\ref{sec:CNM_model}, our numerical simulations are complemented by an analytical model of the warm and cold phase distribution for a specific UV field and ISM mean density. Simulations with a constant UV background provide important comparison cases.

\begin{figure*}
    \includegraphics[width=\textwidth]{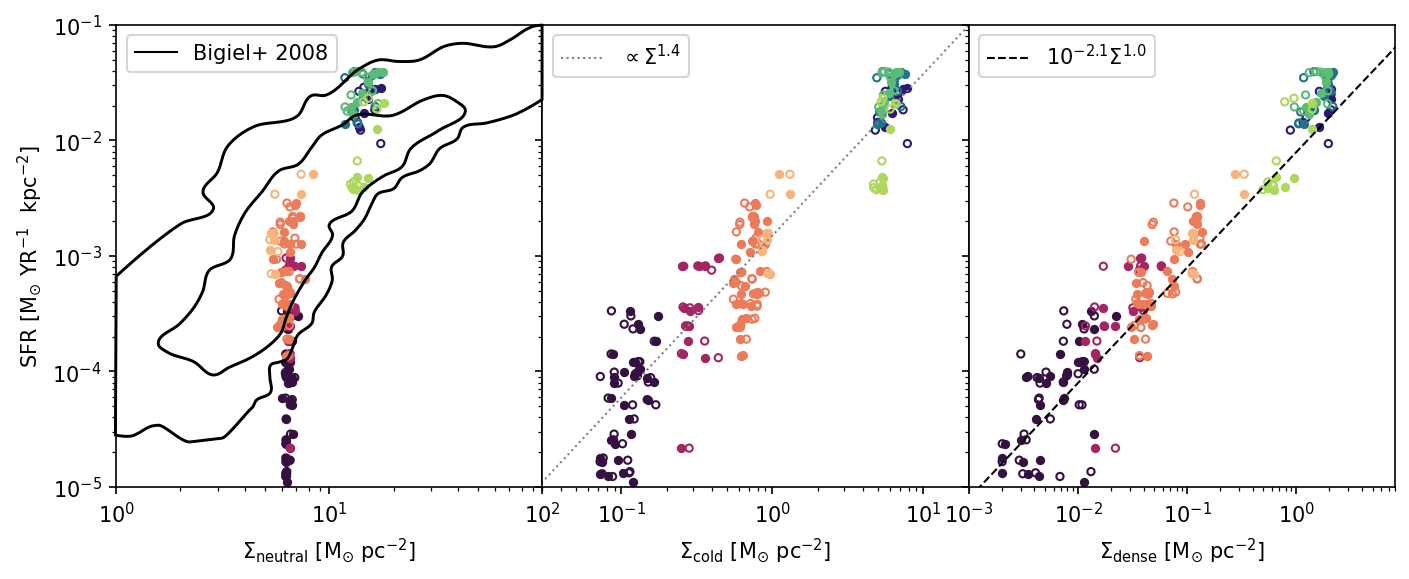}
    \includegraphics[scale=0.75]{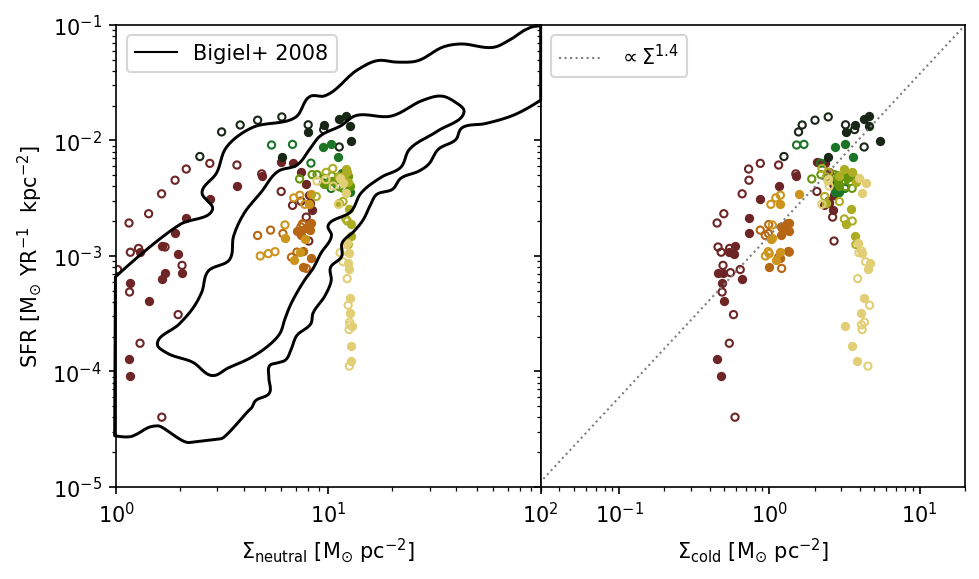}
    \caption{Relations of the SFR with column density for different gas reservoirs.
    Top: simulation set with constant UV (Table~\ref{tab:sims}), bottom: set with variable UV (Table~\ref{tab:sims_UVvar}). The SFR is averaged over a time scale of 10 Myr after (full circles) or before (empty circles) the time of the snapshot at which the gas column densities have been measured. In the left panels, the contours sketch the location of the data point from \citet{Bigiel_et_al2008}: the outer contour tracers roughly all points, while inner one corresponds roughly to their orange contours denoting 2 independent data point per 0.05 dex. In the middle panels, we show a power-law with index 1.4 to guide the eye. In the right panel, we show the linear relation found by \citet{Bigiel_et_al2008} between the SFR and the column density of H$_2$ gas, and extrapolate to lower densities.
    The simulation data is colour-coded as in Figures~\ref{fig:gas_phase_fractions} and~\ref{fig:gas_phase_fractions_ZG}.}
    \label{fig:KS_relation}
\end{figure*}

\section{Star formation relations}
\label{sec:SF_relations}

In Section~\ref{sec:ISM_phases} we have investigated the effect turbulence driving has on the amount of cold gas and dense gas. In this section, we link changes to these mass reservoirs to changes in star formation activity.

\subsection{SFR versus different gas phases}

As discussed in the introduction, observations have revealed correlations between the gas surface density and the SFR.
In Figure~\ref{fig:KS_relation} we correlate the SFR with different gas reservoirs in the simulations.
The SFR is measured 
by averaging 10 Myr after (full circles) or before (empty circles) the time of the snapshot at which the gas column densities have been measured. We verified that our conclusions do not depend on the value chosen for the time-scale of averaging by recomputing the figure for times between 2.5 Myr and 40 Myr. In fact, the correlations presented slightly improve with increasing averaging times.
Observations usually distinguish between HI and H$_2$, the later estimated from CO emission. The simulations with constant UV field do not track chemical evolution and it is non-trivial to estimate the fraction of the neutral gas which would be H$_2$. Instead, we will simply separate the dense gas component, that is cold gas with densities exceeding $n_\mathrm{neutral}>100$ cm$^{-3}$ as in Section~\ref{sec:dense_gas}. 
The correlation between SFR and dense gas is omitted for the simulation with variable UV background (bottom row). This is because the spatial resolution of these simulations is lower and therefore the dense gas is not as well described as in the first set of simulations. 
 
In the left panels of Figure~\ref{fig:KS_relation}, we show the total neutral gas versus the SFR.
We compare with the observational results from \citet{Bigiel_et_al2008}, who studied the KS-relation in a sample of 18 galaxies. Their results for HI + H$_2$ are shown as contours, where the outer contour sketches the full extend of their data points and the inner contour indicate where most of the measurements are situated.
The observations are characterised by a break at 9 \Msun pc$^{-2}$. For higher column densities, a power-law relation with index 1.4 is found, while for lower column densities the relation is steeper with a lot of scatter \citep{Kennicutt&Evans2012}.
The results from the high-$\Sigma$ runs with constant UV field (top panels) fall in the observed range, though the SFR might be a bit high compared to the majority of the observations.
Possibly this is because we do not include early feedback in the form of stellar winds and jets, which are expected to reduce the SFR by a factor 2 \citep{Rathjen_et_al2021, Verliat_et_al2022}.
At low column density, we reproduce the large scatter. The simulations without driving produce SFRs which are below the observed range.
This could mean that the environment modelled by this simulation is unlikely to occur in real galaxies.
Indeed, large-scale dynamic effects such as rotation or infall would provide a minimal level of turbulence driving.
The simulations with strongest driving (orange and green points) generally agree best with the observations.

The behaviour is slightly different for the simulations with variable UV field (bottom panels). They populate additional regions of the diagram.
While many points follow the observations closely, two of the simulations stand out from the others. 
The simulation with an initial column density of 8.4 \Msun\ pc$^{-2}$ without driving exhibits relatively high SFRs for low column densities of neutral gas. Such points were completely absent for the first set of simulations. They correspond to times in the second half the simulation, after a significant portion of the gas has been ionised by the radiative feedback of massive stars, leading to the low columns densities of neutral material. 
The other outlier is the simulation with a column density of 12.7 \Msun\ pc$^{-2}$ and strong driving. For this run, the SFR is lower than what is expected based on the observations during most of its evolution. This possibly indicates that such high levels of turbulence are not representative of real situations with similar column densities.

The middle panels shows the relation with cold gas, that is CNM and cold dense gas which would be molecular.
A positive correlation now appears at low column densities.
These findings are in line with the results from \citet{Smith_at_al2023} who studied the distribution of CNM in their full galaxy simulation.
Overall, the relation between the column density of cold gas and the SFR exhibits a scatter of about one order of magnitude around a power-law with index 1.4.
This is the same index as found for the relation with total gas, above a column density of 20 \Msun\ pc$^{-2}$.
This leads us to the conclusion that the scatter in the observations of the KS-relation at low column densities, as shown in the left panel, is likely due to regional differences in the CNM fraction.
Following this line of thought, it implies that at higher column densities the CNM fraction on a scale of 1 kpc does not vary significantly between regions.
Indeed, in the simulations we find it to be around 40 \% independent of the external driving strength.

Also for the simulations with variable UV background in the bottom panel, we find a similar correlation between the SFR and the cold gas. 
Again, the strong driving run deviates in behaviour: there is a large variation in SFR for similar column densities. The large velocity dispersions in these runs (see Figure~\ref{fig:velocity_dispersion_ZG}), may prevent the formation of sink-forming gas, as discussed in Section~\ref{sec:dense_gas}.

Finally, the right panel shows the relation of the SFR with the dense, cold gas.
As expected, we find a good correlation especially at higher densities. 
We compare with the linear relation between SFR and H$_2$ found by \citet{Bigiel_et_al2008} and extrapolated to lower column densities and find good agreement.
That the agreement extends to the normalisation of the relation may be fortuitous.
The linearity of the relation is however not a coincidence, as many other studies confirm the existence of a linear relation between dense gas and star formation \citep{Lada_et_al2010, Elia_et_al2022}.

\subsection{Knee-density of the KS-relation: two regimes}

The kink in the KS-relation is observed around 9 \Msun pc$^{-2}$.
The findings from our simulations suggest this marks a transition between two regimes of star formation, each with a different dominant limiting process. We propose the following theory.

In the low density regime, we find a correlation between the amount of CNM and the SFR.
In this regime, star formation is regulated predominantly by the formation of cold clouds. This in turn is set by the complex balance of cooling and heating processes in the ISM, which depend on environmental conditions such as the strength of the UV field and the metallicity of the gas.
Additionally, we found that turbulence can boost the CNM fraction in this regime, due to its ability to generate overdensities in which cooling is more efficient.
Low density regions exposed to an external form of turbulence driving would thus have increased CNM fractions and SFRs compared to unperturbed regions.
The increased amount of cold gas gives rise to an increase in the amount of dense gas which, in turn, is directly and linearly proportional to the SFR.
These environmental dependencies for CNM formation lead to large scatter in the KS-relation at low density.

In the high-$\Sigma$ regime, CNM fractions are naturally higher since the densities at which cooling is efficient are reached more easily, leading to an abundance of cold clouds.
Even though we found small variations in the CNM due to excessive turbulence driving, these could not explain the reduction by a factor 4 of the SFR in the run with strong driving.
In this regime, star formation is limited by the formation of dense clumps inside cold clouds rather than the formation of the clouds themselves.
The role of turbulence changed, as excessive turbulence provides additional support against gravitational collapse and reduces the amount of dense gas available for star formation.

From our simulations, it is not clear whether turbulence still plays an important role in the creation of cold clouds at high densities.
We saw that strong external driving keeps a significant amount of the gas in the thermally unstable phase.
Stellar feedback also injects turbulence and so a certain minimum level of turbulence driving can be associated to a SFR.
Together with the additional UV heating of young stars, this may lead to a self-regulation of the CNM-fraction on kpc scales in regions with $\Sigma >$ 9 \Msun pc$^{-2}$.

\subsection{Current models for the break in the KS-relation}
In the literature, we find two main theories for the origin of the break in the KS-relation.

The model of \citet{Ostriker_et_al2010} emphasizes the role played by local thermal and vertical mechanical equilibrium in the galactic disk.
In particular, a transition in the KS relation is predicted around $\simeq 10$ \Msun\ pc$^{-2}$ as a consequence of the gravity being primarily dominated 
by the stellar potential or the self-gravity of the gas (see for instance their Figure~2).

The other idea is that the location of the knee is associated to the phase change from HI to H$_2$, instead of from WMN to CNM as proposed in this work
\citep{Schaye2004, Krumholz_et_al2009b, krumholz2013}.
This idea is supported by the data from observations which reveals that the SFR is linearly proportional to H$_2$. It also relies on the assumption that 
star formation occurs in molecular gas exclusively.

An important difference between the scenario explored in this paper and these previous studies is that a relatively large set of environmental conditions is explored.
As can be seen from Figure~\ref{fig:KS_relation}, this provides a broad variety of situations.
Apart from the break of the mean slope around 10 M$_\odot$ pc$^{-2}$ in the KS diagram, the dispersion of the SFR at lower column densities is very large, almost 3 dex.
Our work suggests that this is possibly a consequence of the 2-phase physics and the sensibility of the WNM/CNM transition to the local UV background as well as the local turbulent driving. 
Since our simulations are not placed in a general, self-consistent, galactic context it is however hard, at this stage, to assess to what extend the variability of the environmental conditions is sufficient to reproduce the observed variability of the KS relations.


Another important aspect that we are not addressing here is the role of the metallicity.
It is now well established that metallicity has a significant role on the HI component in galaxies \citep{wong2013,schruba2018} and does affect the molecular gas content. 
The possible role metallicity has on the SFR is a bit more controversial.
Some studies \citep[e.g.][]{jameson2016} found possible significant dependences of the SFR with metallicities in agreement with the predictions of the models presented by \citet{Ostriker_et_al2010} and \citet{krumholz2013}
Other studies \citep{cormier2014,bothwell2016} found that whereas metallicities has a strong influence on the HI and H$_2$ contents, it has a weak impact on the star formation rate itself. 
This latter conclusion is also met by the numerical simulations of molecular clouds performed by \citet{Glover2012} where it has been found that the star formation rate would vary by a factor of only 2 when the metallicity was changed by a factor of a hundred.

\section{Model for turbulence induced CNM formation}
\label{sec:CNM_model}

The simulation results indicate that turbulence can trigger the cooling of gas initially in the WNM state, and that this may be an important regulator for star formation in the low density regions that are exposed to a significant UV background.
The mechanism behind this is similar as for the balance of turbulence with gravity: turbulence creates overdensities.
Since cooling is efficient in high density gas, overdensities created by shocks cool fast to the CNM state.

While many numerical studies have investigated structure formation in a turbulent multi-phase ISM \citep{Audit_Hennebelle2005, Hennebelle_et_al2007, Walch_et_al2011, Seifried_et_al2011,Saury_et_al2014, Walch_et_al2015, Kim_Ostriker2017,colman24}, an analytical model that described the interaction of turbulence with the thermal state of the gas is still lacking.
Here, we propose such a model with the aim of predicting the CNM fraction.
The general idea is to integrate the mass-weighted density PDF of a turbulent two-phase ISM above a threshold density derived from the temperature-density curve predicted by the cooling/heating model.

\subsection{General assumptions}

We assume that turbulence is the dominant source of density fluctuations and thus the main mechanism for CNM creation. This assumption may not be valid in regions with strong gradients in the gravitational potential.
We do not specify the source of turbulence, which could be for example stellar feedback and/or large-scale galactic processes. The strength of the turbulence is considered to be a free parameter and a characteristic of the environment. It can depend on the level of star formation and whether the region is subject to large-scale galactic processes.

We do not take into account that CNM formation will lead to star formation which is associated with an increase in UV field strength, ionisation and heating \citep{Ostriker_et_al2010}. This could potentially contribute to self-regulating the CNM fraction.

As in the simulations, we only consider the case of solar metallicity gas and solar neighbourhood abundances for the dominant coolants. Other metallicities can be explored by adapting the cooling/heating prescription on which the model is build.

\subsection{Heating/cooling model and threshold density}

\begin{figure}
    \centering
    \includegraphics[width=0.8\columnwidth]{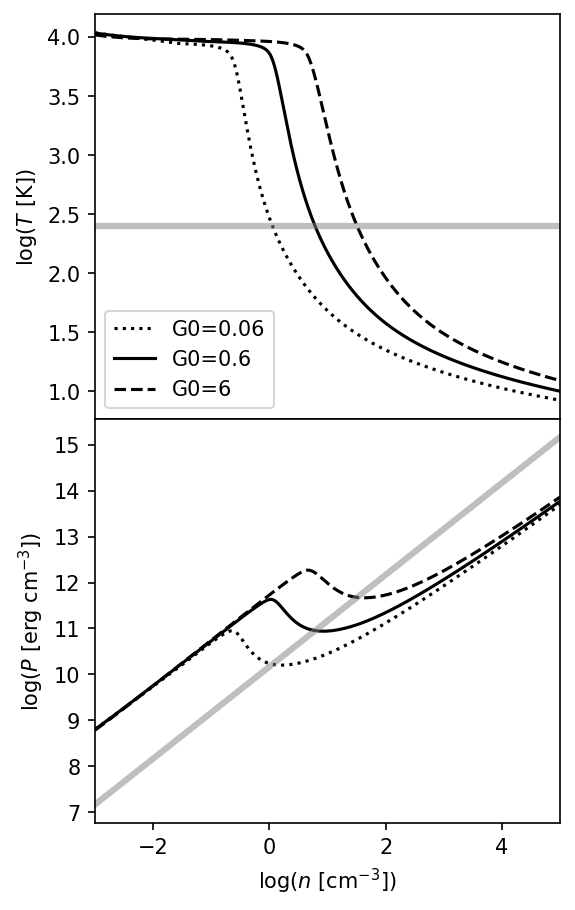}
    \caption{Theoretical temperature-density and pressure-density curve obtained from the cooling model of \citet{Wolfire_et_al2003}, as described in Section~\ref{sec:cooling_heating_model}, for different values of the UV field strength $G_0$. The gray line indicates the chosen temperature threshold for CNM.}
    \label{fig:cooling_model}
\end{figure}

We use the cooling/heating model from \citet{Wolfire_et_al2003} described in Section~\ref{sec:cooling_heating_model}.
The combination of processes leads to the curves shown in Figure~\ref{fig:cooling_model}. 
The only parameter we vary here is the UV field strength $G_0$.
A strong UV field will boost photoelectric heating and move the WNM -- CNM transition to higher densities, while leaving the shape of the curve relatively unchanged.

To determine the threshold density $n_\mathrm{thr}(G_0)$,
we could calculate at which density the sign of the derivative $dP/dn$ changes. The first change, from positive to negative, marks the transition from WNM to LNM. The second change, from negative to positive, is associated with the transition from LNM to CNM.
Here, we simply impose the same temperature thresholds as used to define the phases in the simulation, 3000 K and 250 K, and determine numerically to which density this corresponds.
The 250 K threshold for CNM formation is marked with a gray line in Figure~\ref{fig:cooling_model}. It agrees well with the bend in the pressure-density curve.

\subsection{Density PDF generated by turbulence}

\begin{figure*}
    \centering
    \includegraphics[width=\textwidth]{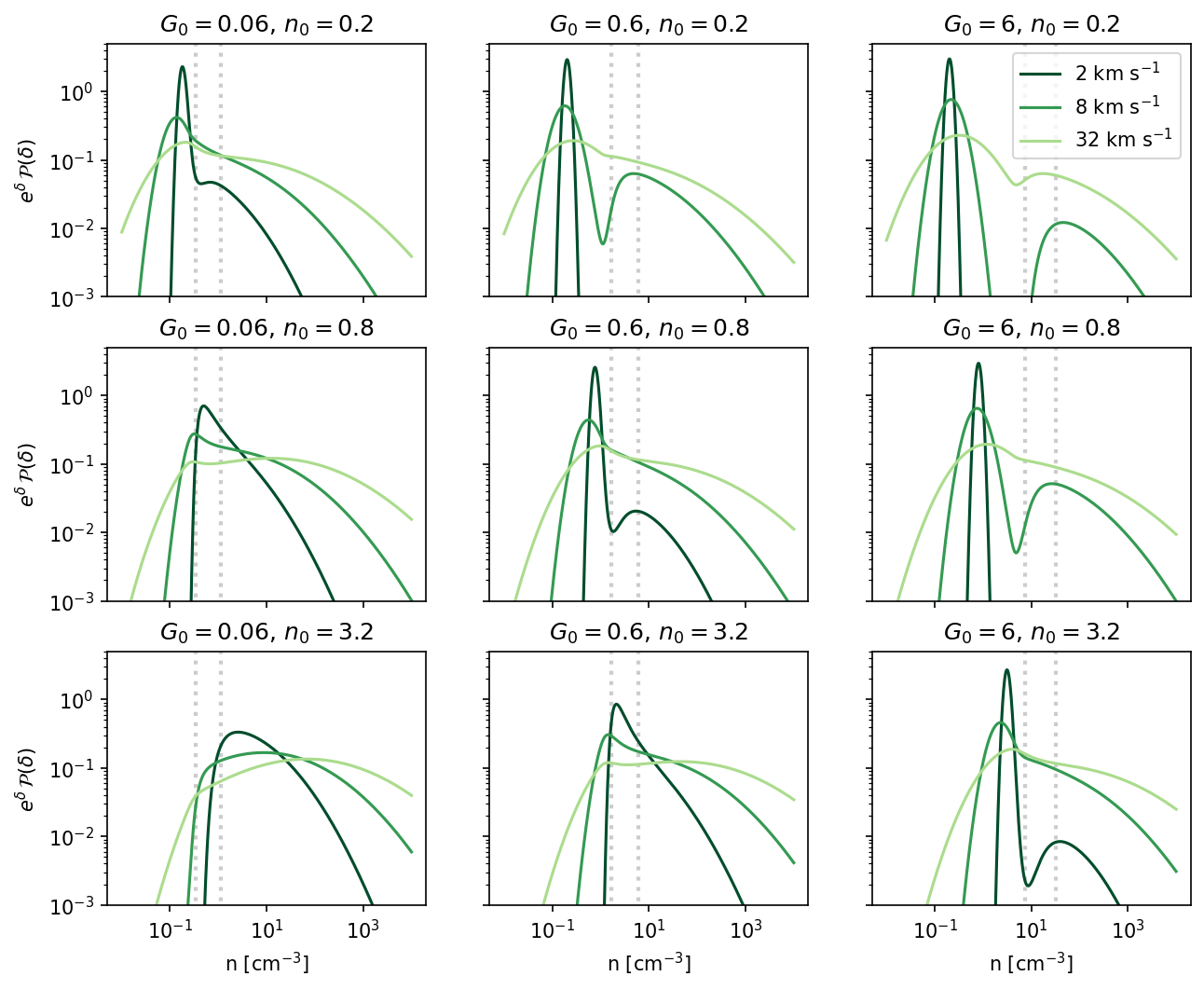}
    \caption{Model for the mass-weighted density PDF of a turbulent two-phase ISM subject to the relation between temperature and density shown in Figure~\ref{fig:cooling_model}. The parameter dependence on UV field strength $G_0$, average density $n_0$ and 3D velocity dispersion $\sigma$ in km/s is explored. The dotted lines mark the densities corresponding to the boundary temperatures, 3000 K and 250 K, used to separate the WNM, LNM and CNM phases.}
    \label{fig:PDF_model}
\end{figure*}

\begin{figure*}
    \centering
    \includegraphics[width=\textwidth]{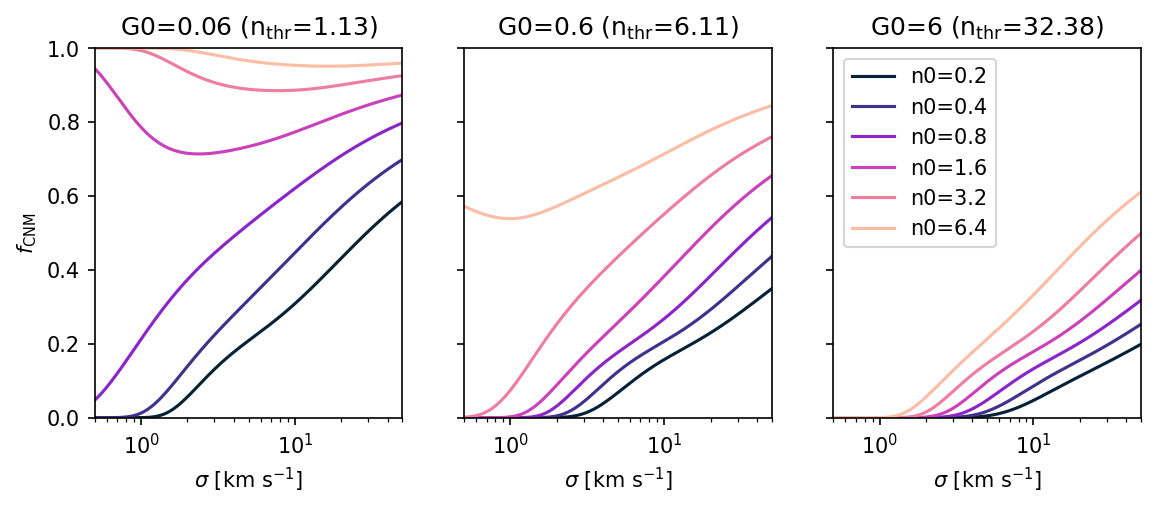}
    \caption{Parameter dependence of the CNM fraction model.}
    \label{fig:fcnm_prediction}
\end{figure*}

It is well-established that turbulence generates log-normal-like density PDFs in isothermal media and that the Mach number controls the width of the distribution \citep{Federrath_et_el2008,federrath10}, as highly turbulent regions contain stronger density fluctuations.
As a function of the density contrast 
$ \delta = \ln(n/ n_0)$, where $n_0$ is the average number density of the region, the density PDF of a turbulent isothermal medium can be described by
\begin{eqnarray}
\label{eq:pdf}
\mathcal{P}(\delta) &=& \dfrac{1}{\sqrt{2 \pi S}} 
\exp\left(- \dfrac{(\delta - \bar{\delta})^2}{2 S}\right)
\end{eqnarray}
with
\begin{eqnarray}
\label{eq:S_old}
S &=& \ln\left(1 + b^2 \mathcal{M}^2\right) \\
\bar{\delta} &=& -S/2,
\end{eqnarray} 
For the compressibility factor $b$, we assume the value of 0.5 for a natural mix of solenoidal and compressive modes.
$\mathcal{M} = \sigma / c_\mathrm{s}$ is the Mach number, with $\sigma$ the 3D velocity dispersion generated by the turbulence and $c_\mathrm{s}$ the sound speed in the medium.

\subsection{Modified density PDF of a two-phase ISM }

The sound speed in the neutral ISM depends on the temperature
\begin{equation}
    c_\mathrm{s} = \sqrt{\frac{k_\mathrm{B} T}{1.4\, m_\mathrm{p}}},
\end{equation}
where 1.4 is the molecular weight of a mixture of atomic hydrogen and helium.
The ISM is however not isothermal.
Due to the temperature difference, the sound speed in the WNM is typically about a factor 10 larger than in the CNM.
As a result, the Mach number of the WNM is smaller than that of the CNM.

Instead of choosing one constant sound speed to calculate the Mach number that regulates the width of the density PDF, we explicitly take into account the dependence of the temperature. 
Using the relation between temperature and density dictated by the heating/cooling model, shown in Figure~\ref{fig:cooling_model}, this temperature dependence transforms into a density dependence.
This effectively makes $S$ density dependent.
Writing this explicitly, we obtain
\begin{eqnarray}
\label{eq:S_new}
S(n) &=& \ln\left(1 + b^2 \frac{1.4 \, m_\mathrm{p} \, \sigma^2}{k_\mathrm{B} T(n)}\right).
\end{eqnarray}
Note that this implicitly also adds a dependence on $G_0$, which we consider constant here. We assume that $\sigma = \sigma_\mathrm{WNM} = \sigma_\mathrm{CNM}$, which may not be the case.

When replacing $S$ by $S(n)$, we are not guaranteed to preserve the normalisation of the lognormal PDF. To solve this issue, we introduce two coefficients $A$ and $B$,
\begin{eqnarray}
\label{eq:pdf_new}
\mathcal{P}(\delta) &=& \dfrac{A}{\sqrt{2 \pi S(\delta)}} 
\exp\left(- \dfrac{(\delta - \frac{S(\delta)}{2} +B)^2}{2 S(\delta)}\right)
\end{eqnarray}
such that 
\begin{eqnarray}
\int \mathcal{P}(\delta) \, \mathrm{d}\delta = 1 \\
\int e^{\delta} \mathcal{P}(\delta) \, \mathrm{d}\delta = 1
\label{eq:normB}
\end{eqnarray}
This ensures that the PDF is well normalized in probability and in mass.

Replacing Eq.~(\ref{eq:S_old}) by Eq.~(\ref{eq:S_new}) results in a density PDF with a more complex dependence on $n$, as shown in Figure~\ref{fig:PDF_model} for a variety of values for the parameters $\sigma$, $n_0$ and $G_0$.
Interestingly, in most configurations we reproduce a bimodal shape similar to the one seen in the density PDF of the simulations (Figure~\ref{fig:phase_diagram}).
This is due to the combination of the low Mach number WNM which produces a narrow log-normal component, with a higher Mach number CNM described by a broader distribution.

The narrow WNM peak is most pronounced when $G_0$ is high or $n_0$ is low. 
Under these conditions, the WNM is the dominant phase.
On the other hand, when $n_0$ is high and $G_0$ low, the broad CNM distribution is dominant and the WNM peak barely noticeable.
Both components broaden with increasing levels of turbulence, now quantified directly through $\sigma$ rather than $\mathcal{M}$.

The relative importance of the broad versus the narrow component sets the CNM fraction.
The model qualitatively captures the behaviour observed when increasing the external driving strenght in the low-$\Sigma$ simulations: when $\sigma$ increases, the WNM peak becomes less pronounces, while the opposite happens for the broad CNM contribution.


\subsection{Parameter dependence of the predicted CNM fraction}

To obtain the CNM fraction $f_\mathrm{CNM}$ in a region of turbulent ISM characterised by $\sigma$, $n_\mathrm{0}$ and $G_0$, we integrate the modified mass-weighted log-normal density PDF above the CNM density threshold:

\begin{equation}
    f_\mathrm{CNM} = \int_{n_\mathrm{thr}(G_0)}^{\infty} e^{\delta} \mathcal{P}(\delta,n_0,\sigma) \mathrm{d}\delta.
\end{equation}
Equation~(\ref{eq:normB}) ensures that the result is properly normalised: $0 \leq f_\mathrm{CNM} \leq 1$.
The resulting CNM fraction and its dependence on $\sigma$, $n_0$ and $G_0$ is illustrated in Figure~\ref{fig:fcnm_prediction}.

The behaviour of $f_\mathrm{CNM}$ with increasing $\sigma$ depends on how the average density compares with the CNM threshold density. 
If $n_0 < n_\mathrm{thr}$, the CNM fraction is zero at low levels of turbulence and increases steadily with $\sigma$.
When $n_0 > n_\mathrm{thr}$, the medium is predicted to be fully in the CNM state at low levels of turbulence. Adding turbulence now typically broadens the density PDF to densities below the CNM threshold, slightly decreasing $f_\mathrm{CNM}$ in favour of the creation of LNM and WNM.
This is a theoretical form of CNM destruction by turbulence which was not clearly observed in the simulations.
The transition occurs around $n_0 \approx n_\mathrm{thr}$, where we find a non-zero minimum in the $f_\mathrm{CNM}(\sigma)$ curve. At very high $\sigma$,
a point of diminishing returns is reached where increasing the turbulence further provides a smaller gain in $f_\mathrm{CNM}$.
This is because the PDF broadens not only to high densities but also to low densities as large voids are being created as counterparts to the dense shocks.


\subsection{Comparison to simulation results}

\begin{figure*}
    \centering
    \includegraphics[width=\textwidth]{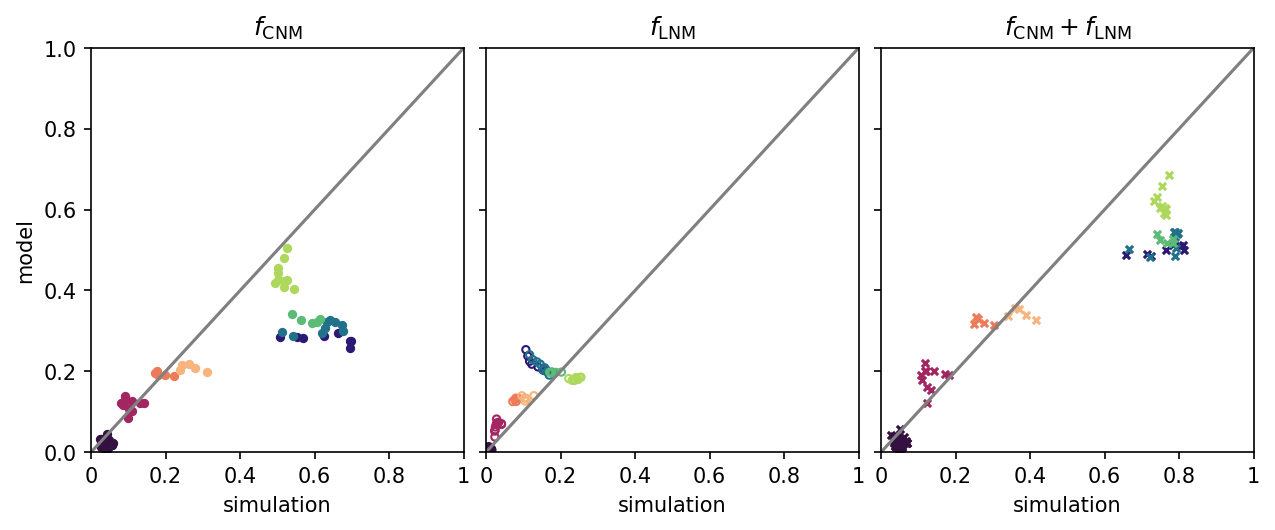}
    \caption{Comparison between the mass fractions measured in the simulation with the ones predicted by the model. We show the fraction of CNM (left), LNM (middle) and the sum of both, which is equivalent to $1 - f_\mathrm{WNM}$.}
    \label{fig:model_vs_sim}
\end{figure*}

To test the accuracy of our model, we want to quantitatively compare with the simulation results.
One complication is that galactic disks are not uniform but on average have a Gaussian density profile around a mid-plane density.
To partially mitigate this issue, we consider only the inner region of the disk, that is a band between $z_\mathrm{midplane} - 100$~pc and $z_\mathrm{midplane} + 100$~pc.

For each snapshot, we measure the average total density of neutral gas $n_0$ and the velocity dispersion $\sigma$ in the mid-plane region to input into our model for $f_\mathrm{CNM}$.
The average density varies over time as the vertical balance in the disk first evolves from the initial conditions to balance the various supporting pressure forces against gravity, and later on adapts to the varying levels of star formation activity.
We find values between 0.35 and 1.55 cm$^{-3}$ across the simulations in the set with constant UV background.
The velocity dispersion results from the combination of external turbulence driving on large scales and stellar feedback.
We find values between 2 and 25 km s$^{-1}$.
We set $G_0=0.6$, as this is the value of the constant UV background used in the cooling routines.


\subsubsection{Gas phase fractions}

In Figure~\ref{fig:model_vs_sim}, we compare the model predicted phase fraction to the ones measured in the simulations with constant UV background\footnote{Note that the measured CNM fractions here are higher than those shown in Figure~\ref{fig:gas_phase_fractions} since here we are limiting ourselves to the densest part of the disk.}.
While our main goal is to model the amount of CNM, we also verify what the model predicts for the LNM.
For the low-$\Sigma$ runs, which have fairly low $f_\mathrm{CNM}$, the prediction is excellent (left panel, bottom left corner). The model accurately describes the trend of increasing $f_\mathrm{CNM}$ with increasing external turbulence driving strength.
The agreement is less satisfying for the high-$\Sigma$ simulations.
While $f_\mathrm{CNM}$ is predicted correctly in the strong driving case, it is underestimated in the other three simulations.
The discrepancy can be reduced by increasing the parameter $b$ of Eq.~\ref{eq:S_new} for these three runs.
This is justified by their higher SFR which lead to more SN feedback, a mostly compressive turbulence driver.
Gravitational compression is also stronger when the average density is higher.

The prediction for the LNM fraction is shown in the middle panel.
In most cases, it is slightly overestimated and for the high-$\Sigma$ runs we do not reproduce the expected increase with turbulence driving strength. 
Our model does not take into account the unstable nature of the LNM, which could explain why we overestimate it.

Looking at the right panel, we see that the agreement between the model and the simulations for the sum $f_\mathrm{CNM} + f_\mathrm{LNM}$ (or equivalently $f_\mathrm{WNM}$ is good for low column density. For the high-$\Sigma$ runs, the WNM fraction is overestimated by the model.

\subsubsection{Density PDF}

\begin{figure*}
    \centering
    \includegraphics[width=\textwidth]{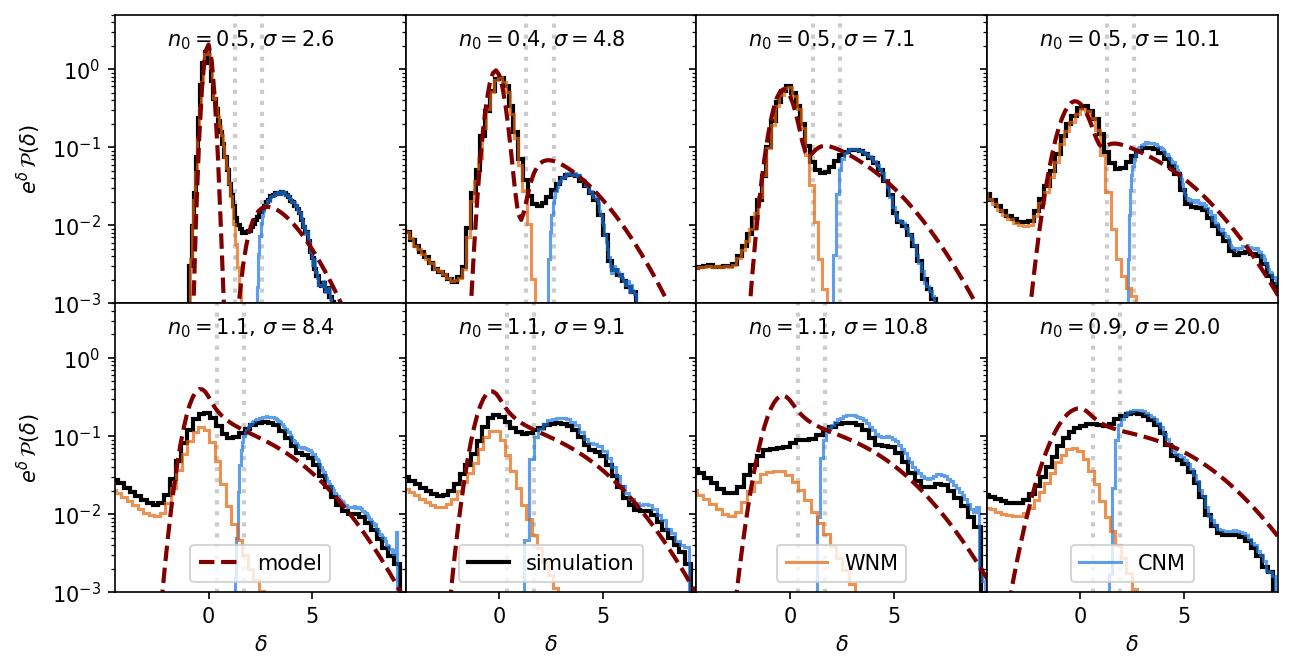}
    \caption{Comparison between the measured mass-weighted density PDF and the modelled PDF for a representative snapshot of each simulation.
    The black curve corresponds to the total neutral gas in the simulation, while the orange and blue curves show the contribution from the WNM and CNM respectively.
    The dashed line shows the model calculated using the average density and velocity dispersion measured in the snapshot.
    Only the inner part of the disk, between -100 and +100 pc around the mid-plane, is considered for the measurement of the PDF and average properties.
    The low-$\Sigma$ runs are shown in the top panels, the high-$\Sigma$ ones in the bottom. External driving strength increases from left to right. 
    }
    \label{fig:PDF_sim_vs_model}
\end{figure*}

We can gain more insight in the results by directly comparing the measured density PDF with the modelled one.
In Figure~\ref{fig:PDF_sim_vs_model}, we show such a comparison for a selection of example snapshots, one for each simulation.
Generally, the WNM peak around 1 cm$^{-3}$ is quite well-described by the model.
The point of transition to CNM is also captured reasonably well.
What does not match well is the shape of the CNM component.
In most cases, the model does not reproduce the position of the peak of the distribution.
This component of the PDF is also too broad in some cases, resulting in an overestimation of the amount of \textit{dense} CNM.
Even though the shape of the CNM component leaves much to be desired, the relative importance of CNM versus WNM is well-captured, resulting in a good prediction for $f_\mathrm{CNM}$.

\subsection{Caveats and further improvements}

Modelling the density PDF of the multi-phase ISM is a difficult problem.
Here we proposed a basic model with the aim of self-consistently estimating the CNM fraction.
In this, the model succeeds surprisingly well, given that all we did was introduce a density-dependent sound speed.
However, more work is needed to accurately describe the full shape of the density PDF.
While this is beyond the scope of this paper, we list here some of the ideas we started exploring.

A first avenue is to take into account the magnetic field. 
\citet{Molina_et_al2012} found that this can be achieved by including an additional factor in $S$:
\begin{equation}
    S = \ln \left( 1 + b^2 \mathcal{M}^2 \frac{\beta}{(1 + \beta)} \right)
\end{equation}
where $\beta = P_\mathrm{thermal} / P_\mathrm{magnetic}$.
When $\beta<1$, this would reduce the width.
This dependence is not unlike the dependence on $b$, which decreases the width when the turbulence is more solenoidal.
We tested several values of $b$ and found that assuming the turbulence is fully solenoidal ($b=0.3$) decreases the amount of dense gas while only slightly reducing the CNM prediction. However, this did not solve the problem of the mismatched CNM peak.

It has been pointed out that the PDF may not be exactly log-normal, but has a skewness to it especially at high Mach numbers \citep{Hopkins2013, Hennebelle_et_al2024, Brucy_et_al2024}.
This may be important in the CNM. For the low Mach numbers that characterise the WNM, the log-normal form is a good approximation.
We tried replacing the log-normal by this skewed version, dubbed the Castaing-Hopkins PDF \citep{Castaing1996, Hopkins2013}.
While this produces a better approximation for the transition to LNM and an improved peak shape for the CNM, it introduced a sharp cut-off at high densities and essentially predicted zero gas with densities above 100 cm$^{-3}$.
One could argue that at this point gravity would significantly alter the density distribution, producing the additional power-law-like tail \citep{Burkhart_et_al2017}.

The lack of high density gas was not the only issue with the Castaing-Hopkins PDF. While it can reproduce the shape of the WNM component when modelling the high-$\Sigma$ simulations, for the low-$\Sigma$ runs the WNM peak is shifted too far to the right.
These experiments indicate that the WNM is better described by a log-normal, while the CNM may consist of a Castaing-Hopkins PDF with an added power-law to include gravity.

That we do not capture the characteristic LNM minimum and CNM maximum, is likely due to the fact that we do not actually take into account that the LNM is an unstable equilibrium.
The gas is more likely to be in one of the stable states, that is WNM or CNM.
Furthermore, gas that is in the CNM and WNM has a preferred pressure and thus density. This pressure is given by the various vertical balancing forces in a galactic disk, among which feedback from star formation \citep{Ostriker_et_al2010, Ostriker_Kim2022}. 
Indeed, in Figure~\ref{fig:phase_diagram} we see the distribution of pressures is broad but peaked.
This should somehow be taken into account.

An alternative approach would be to model the WNM and CNM separately and sum the two to obtain the total PDF. Each PDF would be centred around a characteristic density obtained from a pressure equilibrium model. The width is obtained using the corresponding temperature and optionally a density-dependent velocity dispersion. 
The problem with this approach is that the CNM and WNM fractions need to be assumed or modelled separately, since PDF = $f_\mathrm{WNM}$ PDF$_\mathrm{WNM}$ + $f_\mathrm{CNM}$ PDF$_\mathrm{CNM}$.
Since our base model gives good results for the CNM fraction, a possibility could be to use this to normalise the individual contributions.

\section{Summary and conclusions}
\label{sec:summary}

In this study, we investigate the effect of large-scale turbulence driving on star formation and the thermal balance of the ISM.
We analyse a suite of numerical simulations of a 1 kpc$^3$ part of a galaxy including stellar feedback in the form of SN and HII regions. The cooling model reproduces a multi-phase ISM with warm and cold neutral phase.
The simulation suite explores the star formation behaviour for several initial column density values representative of regions in the Milky Way ranging from 8.4  to 19 \Msun pc$^{-2}$ .
For each subset of simulations, we tested different strengths of the external large-scale turbulence driving, including a reference simulation without any external driving.
Regarding the UV background, two assumptions are considered, either a constant value or a time-dependent one, proportional to the star formation rate in the 
computational box.

We found that the additional turbulence is not only capable of decreasing the SFR, as found in previous studies, but can also enhance it under the right conditions.
In a medium with low column density exposed to a constant UV background the SFR increases with increasing external driving strength and the star formation activity is easily an order of magnitude higher than in the case without driving. When the column density was increased, we saw that when the external LS driving becomes significantly stronger than the turbulence injection from SN, the SFR is reduced.
Also in the case of an SFR-dependent UV background, additional turbulence queches star formation.

The changes in SFR are linked to the thermal state of the gas.
In the lower column density simulations with constant UV background, increasing the level of turbulence boosts CNM formation which in turn leads to more dense gas for star formation. The CNM fraction increases with driving strength.
For the higher column density simulations, the CNM fraction is similar for all runs, while the LNM fraction increases with increasing levels of turbulence as it acts as a perturbing force and promotes phase change.
Additionally, the amount of \textit{dense} CNM is reduced, leading to a smaller star formation rate.
In the simulations with a time-dependent UV background, the CNM fraction is reduced when additional external turbulence is present.

Studying the relation between the SFR and the column density of different mass reservoirs, we further characterise the two regimes.
Overall we find a linear relation between the dense gas and the SFR, as in observations. We identify an additional correlation between the SFR and the CNM. This is particularly striking for the low column density runs with constant UV background, were no correlation is found between the SFR and the full mass reservoir of neutral gas (including WNM).
Our interpretation is that at low densities or high UV backgrounds, star formation is limited by the formation of cold gas clouds.
At higher densities or low UV backgrounds, an abundance of cold clouds exist and the limiting process is the formation of dense clumps inside those clouds.
This change in regime may be an explanation for the observed knee in the Kennicutt-Schmidt relation. This would imply that the scatter in this relation at low column densities is due to variations in the CNM fraction.

To better understand the mechanism behind the generation of CNM by turbulence, we introduce an analytical model containing two main ingredients: the density PDF generated by turbulence and the theoretical temperature-density curve given by the ISM cooling/heating model.
To calculate the Mach number, which controls the width of the density PDF, we use a density-dependent sound speed derived from the $T - \rho$ curve.
This leads to a bimodal shape for the PDF, with a narrow WNM peak and a broad CNM component.
Their relative importance then sets the CNM fraction, which we compute by integrating the modified PDF above the density associated with the maximum CNM temperature.

Our analytical model predicts that when the average density of a region is low compared to the threshold density for CNM formation (which is dependent on the UV background), increasing the level of turbulence will boost the CNM fraction.
This is because more shocks and sharper density contrasts are created at higher Mach numbers, resulting in a widening of the density PDF which now extend to values above the CNM threshold.
On the other hand, if the average density was already larger than the CNM threshold, all the gas will be in the cold phase at low levels of turbulence. Increasing the turbulent velocity dispersion broadens the PDF both to higher and lower densities, resulting in the generation of WNM and a small decrease in CNM fraction.

Using this model, we are able to predict the CNM fraction for most of the simulations {with a constant UV background}. The match is especially good in the low column density regime where the WNM is the dominant phase.
However, when directly comparing the model density PDF to the one measured in the simulations, we see that the shape of the CNM component is not well-described.
This shows that more work is needed to accurately model the complex ISM.

To conclude, in general turbulence quenches star formation. But when the heating-cooling balance is in favour of forming WNM rather than CNM, which can be the case when the region is subject to an external heating source or lacks the metals that provide the main ISM cooling, turbulence can boost star formation by providing overdensities in which the ISM can cool more easily. It remains to be investigated, using observational data, where these types of environmental conditions occur in the real ISM, and whether this can indeed be linked to the knee in the Kennicutt-Schmidt relation.

\section*{Acknowledgements}
We thank the anonymous reviewers for their useful comments that led to significant improvements to this work. 
This research has received funding from the European Research Council
synergy grant ECOGAL (Grant  855130).
TC would like to thank all the members of the ECOGAL collaboration for discussions and comments which greatly improved this work.
The authors acknowledge Interstellar Institute's programs "II4: The Grand Cascade",
as well as the Paris-Saclay University's Institut Pascal for hosting discussions 
that triggered this work and nourished the development of the ideas behind it.
This work was granted access to HPC resources of CINES and CCRT under the allocation x2020047023 made by GENCI (Grand Equipement National de Calcul Intensif).
NB acknowledges support from the ANR BRIDGES grant (ANR-23-CE31-0005).
In addition to ECOGAL, RSK and SCOG acknowledge funding from the German Excellence Strategy via the Heidelberg Excellence Cluster STRUCTURES  (EXC-2181/1 - 390900948).
TC acknowledges funding received from the European High Per-
formance Computing Joint Undertaking (JU) and Belgium, Czech
Republic, France, Germany, Greece, Italy, Norway, and Spain under
grant agreement no. 101093441 (SPACE).

\section*{Data Availability}
The simulations will be made available on the Galactica database
under the tag \texttt{LS\_DRIVING}.



\bibliographystyle{mnras}
\bibliography{large_scale}



\appendix

\section{WNM temperatures for different ionisation fractions}

\begin{figure}
    \centering
    \includegraphics[width=0.48\columnwidth]{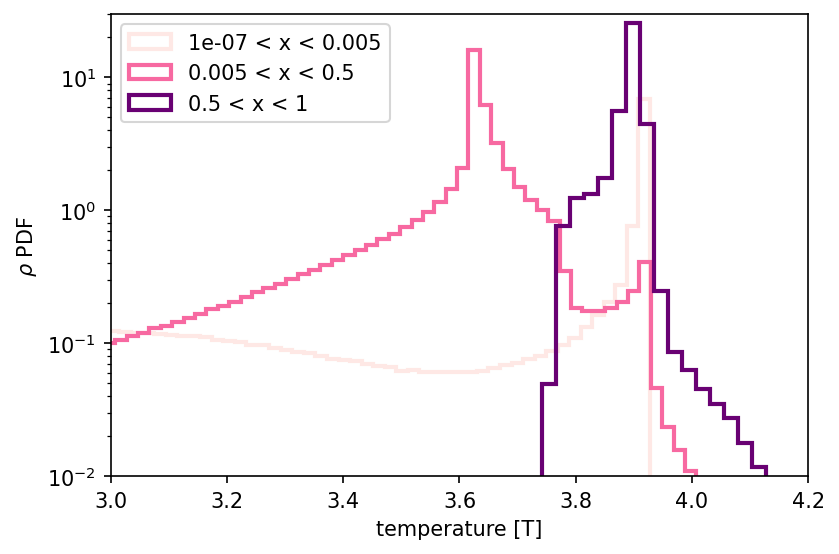}
    \caption{Mass weighted temperature PDF for different levels of ionisation fraction $x$ for an example snapshot.}
    \label{fig:pdf_temperature_xion}
\end{figure}

Figure~\ref{fig:pdf_temperature_xion} shows a mass weighted histogram of the WNM temperatures in an example snapshot of our simulations. The histogram is split according to the ionisation fraction of the gas $x$. We see that gas with an ionisation fraction between 0.5 and 50\% has a lower temperature than gas which is more than 50\% ionised. This results in two distinct branches of WNM in the phase diagram in Figure~\ref{fig:phase_diagram}. This can likely be associated to inconsistent way of calculating the ionisation for the cooling model. fractions. in the treatment of cooling and ionisation, as mentioned in Section~\ref{sec:cooling_heating_model}

\section{The influence of the supernova treatment}
\label{sn_treatment}

\begin{figure}
    \centering
    \includegraphics[width=0.48\columnwidth]{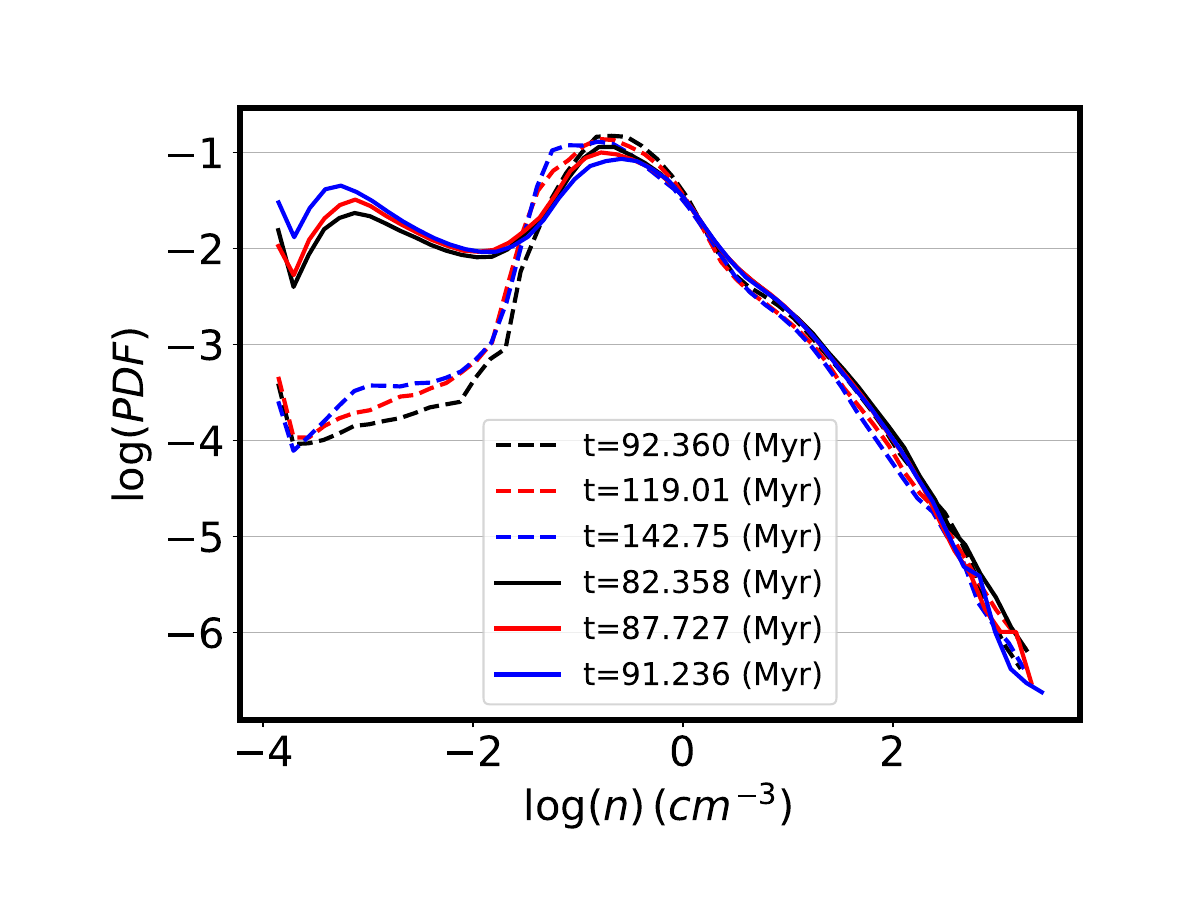}
    \includegraphics[width=0.48\columnwidth]{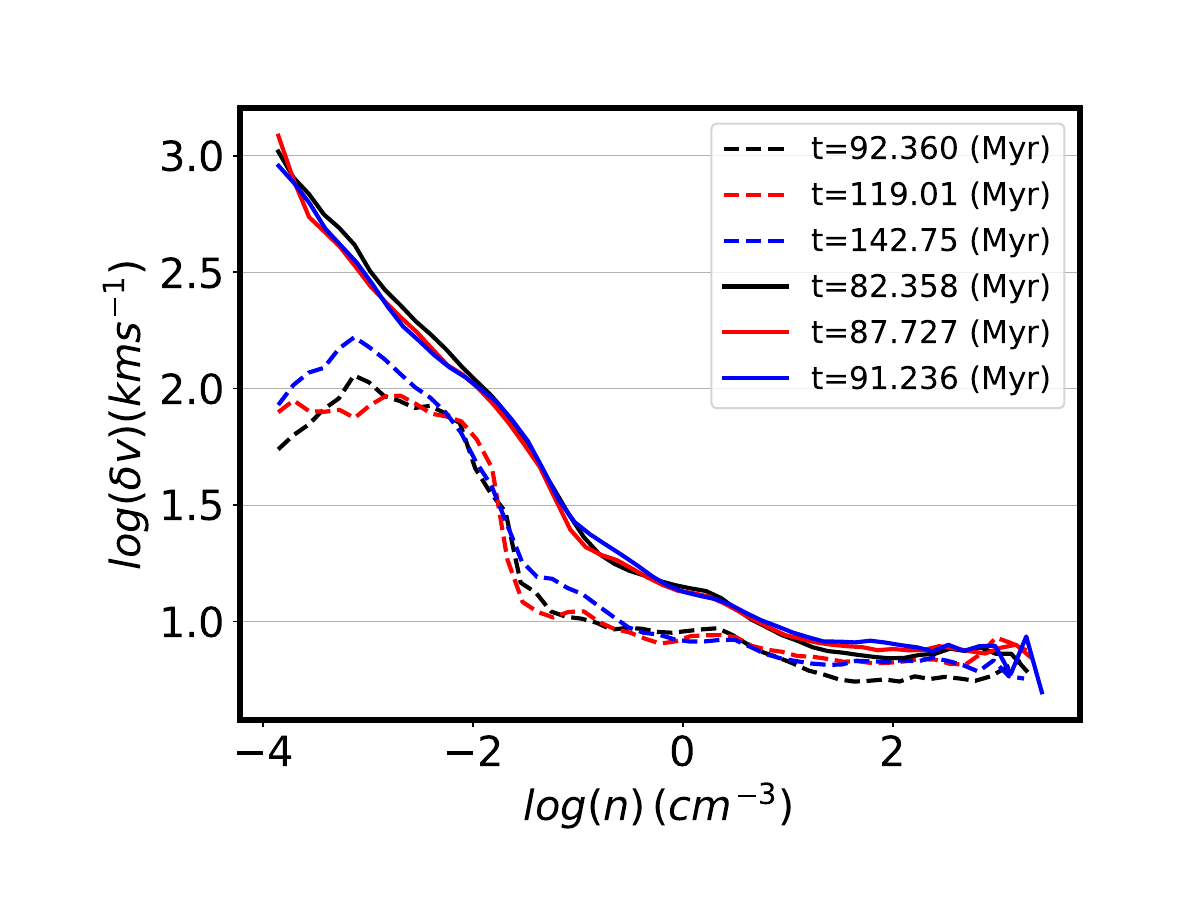}
    \caption{Comparisons between simulations for which the supernova velocities are limited to 300 km s$^{-1}$ (dashed lines) and simulations where they are 
    limited to 3000 km s$^{-1}$ (solid lines). The left panel portrays the volume weighted density PDF and the right panel the 
    mean velocity in density bins.}
    \label{fig:SN_treatment}
\end{figure}

As explained above, the time step induced by supernova explosions is rather short. 
For instance, with a resolution of 4 pc it is about 100 yr or less, making the simulations very timely, even with a 
relatively coarse spatial resolution.
For a sub-parsec resolution, as we have in this work, this would produce time steps on the order of 10 yr or even less. To circumvent this difficulty, we have limited the velocity of the supernova remnant to 300
km s$^{-1}$ and the temperature of the gas to 10$^6$ K.
With these limits, the time steps is about 10 times larger. However, it is important to check the effects that this procedure may have on the results.
Figure~\ref{fig:SN_treatment} compares the results of simulations where the velocities and temperatures use the limitations listed in Section~\ref{sec:setup}
with those of simulations for which the limits are set to much higher values (3000 km s$^{-1}$ and 10$^8$ K). On the left, we show the density PDF and on the right is the velocity dispersion within density bins.
The column density of these two simulations is 19 M$_\odot$ pc$^{-2}$. 
As can be seen, large differences are found for the diffuse gas (i.e. below ~0.1 cm$^{-3}$), for both the density PDF and the velocity dispersion. 
However, the differences for gas denser than ~0.1 cm$^{-3}$ appear to be quite limited.
Let us recall that for the spatial scales at which ISM simulations are run, large velocities appear only when supernovae explode in very diffuse gas (i.e. below 0.1 cm$^{-3}$).
In this case, the supernova explosion has little impact on the dense gas.

\end{document}